\documentclass[fleqn,usenatbib]{mnras}

\usepackage[T1]{fontenc}
\usepackage{ae,aecompl}

\usepackage[pdftex]{graphicx}	% Including figure files
\usepackage{amsmath}	% Advanced maths commands
\usepackage{amssymb}	% Extra maths symbols
\usepackage{newtxtext,newtxmath}
\title[Lifetime with non-equal masses]{Stable lifetime of compact, evenly-spaced planetary systems with non-equal masses}

\author[D. R. Rice et al.]{
David R. Rice,$^{1,2}$\thanks{E-mail: david.rice@unlv.edu}
and Jason H. Steffen$^{1,2}$
\\
$^{1}$Department of Physics \& Astronomy, University of Nevada, Las Vegas, 4505 S. Maryland Pkwy., Las Vegas, NV 89154, USA\\
$^{2}$Nevada Center for Astrophysics, University of Nevada, Las Vegas, 4505 South Maryland Parkway, Las Vegas, NV 89154, USA
}

\date{Accepted 1 February 2023. Received 1 February 2023; in original form}

\pubyear{2019}

\begin{document}
\label{firstpage}
\pagerange{\pageref{firstpage}--\pageref{lastpage}}
\maketitle

\begin{abstract}
Compact planetary systems with more than two planets can undergo orbital crossings from planet-planet perturbations.  The time which the system remains stable without orbital crossings has an exponential dependence on the initial orbital separations in units of mutual Hill radii.  However when a multi-planet system has period ratios near mean-motion resonances, its stability time differs from the time determined by planet separation.  This difference can be up to an order of magnitude when systems are set up with chains of equal period ratios.  We use numerical simulations to describe the stability time relationship in non-resonant systems with equal separations but non-equal masses which breaks the chains of equal period ratios.  We find a deviation of 30 per cent in the masses of Earth-mass planets creates a large enough deviation in the period ratios where the average stability time of a given spacing can be predicted by the stability time relationship.  The mass deviation where structure from equal period ratios is erased increases with planet mass but does not depend on planet multiplicity.  With a large enough mass deviation, the distribution of stability time at a given spacing is much wider than in equal-mass systems where the distribution narrows due to period commensurabilities.  We find the stability time distribution is heteroscedastic with spacing---the deviation in stability time for a given spacing increases with said spacing.  
\end{abstract}


%%%%%%%%%%%%%%%%%%%%%%%%%%%%%%%%%%%%%%%%%%%%%%%%%%

%%%%%%%%%%%%%%%%% BODY OF PAPER %%%%%%%%%%%%%%%%%%

\section{Introduction}\label{sec:intro}

The dynamical spacing of a system influences how a post-gas disc system evolves through planet-planet interactions \citep{fang, volk}.  In a closely-packed system, the planets' eccentricities grow from planet-planet perturbations until an orbital crossing occurs.  A crossing can be marked by a close encounter of two planets after which the system experiences chaotic orbital evolution until a collision or ejection occurs.  The lifetime of the stable system before the instability occurs is dependent on the initial spacing between the planets.

Within a critical spacing, a system's orbits are always chaotic because of overlapping two-body resonances \citep{deck}.  On the other hand, many observed systems are in resonant chains where multiple bodies are in two- and three-body resonances \citep{Mills2016,MacDonald2016,Luger2017,MacDonald2022}.  These resonances stabilize the system \citep{Matsumoto2012,Pichierri2020}.  However, the typical planetary system found by the \textit{Kepler} Space Telescope, even relatively compact systems, are not in resonance \citep{fabrycky,steffen}.  A planetary system's dynamical spacing and closeness to resonance are important considerations in assessing its long term stability. 

\citet{obertas} studies the relationship between stability time and dynamical spacing measured in mutual Hill radii (Eq.~\ref{eq:hill}) with over 10,000 simulations of non-resonant systems.  They use systems of five Earth-mass planets on co-planar and circular orbits with a solar-mass star.  As described in previous work starting with \citet{chambers}, they find a strong, exponential relationship where the stability time increases by approximately an order of magnitude with every integer increase in mutual Hill radii spacing between planets.  

The use of equal-mass bodies with equal Hill separations implies that the bodies are in equal period ratios.  This artificial setup which doesn't replicate observed systems with variable masses places the planets with equal period ratios which cause perturbations at regular intervals between the bodies.  In \citet{obertas}, systems with adjacent or next-adjacent planets near period commensurability with first or second-order mean-motion resonances (MMR) have stability times that deviate from the time predicted by the spacing.  Also reported in \citet{chambers, marzari, pu}, the chains of period ratios near the ratio of a MMR in these multi-planet systems cause separation-dependent modulations superimposed on the stability time relationship that have amplitudes as large as an order of magnitude.   

\citet{chambers} integrates 120 systems of 20 planets with equal spacing and masses that vary by a factor of five.  In one statement, they report the structure from modulations disappearing in these systems.  Here, we investigate with much higher resolution the use of variable masses in more-restricted 4 planet systems with equal Hill spacing.  We detail the amount of mass variation and its corresponding period ratio variation needed to smooth out the stability time relationship.  We detail how inhomogeneities in a system's masses widen the distribution of stability time resulting in a heteroscedastic stability time relationship with spacing. 

Our paper is laid out as follows.  In Section \ref{sec:spacing} and Section \ref{sec:method}, we describe the relevant parameters to our study and the setup of our planetary systems.  We reproduce the resonant deviations in stability time for systems with equal-mass and equal-spacing in Section \ref{sec:stability}.  We then model systems with equal-spacing but use variable mass planets in Section \ref{sec:mass} and explore the amount of variance in mass needed to erase modulations in the stability time relationship for various masses and multiplicities.  Finally, in Section \ref{sec:hetero} we use planets from 1-10 M$_\oplus$ in each system and detail the heteroscedastic relationship of stability time for this large range of planet masses.  We conclude with a summary and discussion in Section \ref{sec:conc}.

\section{Dynamical Spacing and Stability: Relevant Parameters}\label{sec:spacing}

For small eccentricities and inclinations, the mutual Hill radius of two bodies is defined as
\begin{equation}
R_{H_{1,2}} = \left(\frac{m_1 + m_2}{3M}\right)^{1/3}\frac{a_1 + a_2}{2}, \label{eq:hill}
\end{equation}
where $m_1$ and $m_2$ are the planetary masses, $a_1$ and $a_2$ are their semi-major axes, and $M$ is the mass of the central body.  The separation of two planet's semi-major axes can be measured in units of mutual Hill radii by a spacing parameter, $\Delta$, given as
\begin{equation}
a_2-a_1=\Delta R_{H_{1,2}}.\label{eq:sep}
\end{equation}
The spacing is denoted as ``$k$'' in \citet{zhou} and ``$\beta$' in \citet{smith}. 

For a system of \textit{n} planets, Eq.~\ref{eq:hill} and Eq.~\ref{eq:sep} can combine to give 
\begin{equation}\label{eq:space}
a_{i+1} = a_{i}\frac{2 + \Delta K}{2 - \Delta K},
\end{equation} 
where 
\begin{equation}\label{eq:K}
K=\left(\frac{m_i+m_{i+1}}{3M}\right)^{1/3}\
\end{equation} 
for a pair of planets of masses $m_i$ (inner) and $m_{i+1}$ (outer) around a central body of mass $M$.  Once the innermost planet's orbit (starting from $a_1$) and a spacing parameter is chosen, the masses of adjacent planets then dictate the semi-major axis of the next planet. 

A two-planet system is stable if $a_2/a_1$>$1+2\sqrt{3}K$ to the lowest order in planet masses \citep{gladman}.  With Eq.~\ref{eq:space}, the critical spacing parameter for two planet stability is given by 
\begin{equation}\label{eq:delc}
\Delta_c=\frac{2\sqrt{3}}{1+\sqrt{3}K}.
\end{equation} 
However, for systems of more than two planets the energy and angular momentum of a given planet pair are not conserved because of perturbations from the additional planets.  This results in orbit crossings even in systems with initially large separations.

\citet{chambers} find an exponential relationship between the dynamical spacing in multiple planet systems and the time from initial conditions to the first close encounter (defined as a separation of less than one Hill Radius).  We refer to this time as the ``close encounter time'' or ``stability time'' and denote it as $t_c$.  The relationship is given by
\begin{equation}
\log(t_c)= b(\Delta) + c.\label{eq:chamber}
\end{equation}
The values of the constants $b$ and $c$ depend on planet mass and multiplicity \citep{chambers} and initial eccentricity and inclination \citep{yoshinaga}.  To compare worth previous work, we use the mutual Hill radius to measure separations, $\Delta$, though the relationship becomes independent of planet mass when separations are measured in units of $(m_p/m_o)^{1/4}$

Close encounters always happen on the same timescale as orbital crossings.  However, we note that \citet{Rice18} and \citet{Bartram2021} show that the architecture of the system may not change through a collision or ejection of a planet on this same timescale in non-coplanar systems.  The stability or survival time relationship has been detailed in works such as \cite{funk, zhou, quillen, yalinewich,Petit2020,Hussain2020,Lissauer2021} and in the specific case of binary star systems in \citet{quarles}.

For multiple planet systems with equal-mass planets, the equal spacing given by Eq.~\ref{eq:space} keeps the ratio of semi-major axis constant.  From Kepler's Third Law, the period ratio is also constant, 
\begin{equation}\label{eq:pratio} 
P_r = \frac{P_{i+1}}{P_{i}} = \left(\frac{2+\Delta K}{2-\Delta K}\right)^{3/2}.
\end{equation}
In equal-mass, equal-spacing systems the period ratio will be equal for all pairs of adjacent planets.  Additionally, the period ratios of next-adjacent planets (i.e. P$_3$/P$_1$ and P$_4$/P$_2$) will be equal.  

If the period ratio of two planets is equal to a ratio of small integers the planets are said to be near MMR.  When planets in equal-mass, equal-spacing systems are near first and second order MMR the stability time of the system can be an order of magnitude longer or shorter than expected from the exponential Eq.~\ref{eq:chamber} \citep{obertas}.  One other work, \citet{gratia}, shows these modulations in stability time are dampened when initial eccentricities are above 0.05.  In this work, using variable masses changes $K$ in Eq.~\ref{eq:pratio}, breaking the chain of equal period ratio between the planets.

\section{Simulations}\label{sec:method}

To run our N-body integrations we use the orbital dynamics package MERCURY6.2 \citep{chambers2}.  We use the Bulirsch-Stoer method and the accuracy parameter is set to $10^{-12}$.  The initial timestep is set to less than 1/20\textsuperscript{th} of the initial period of the innermost planet.  The planets are placed around a central body of 1.0 $M_\odot$.

The initial conditions of the planets are detailed in Table \ref{tab:IC}.  Our artificial planetary systems in this study will consists of four planets.  We use a Rayleigh distribution to draw eccentricities and inclinations \citep{fang}.  However, we use a small Rayleigh scale parameter of 10$^{-6}$ which keeps the system initially near-circular and near-coplanar to be comparable with \citet{Rice18}.  The three remaining orbital parameters---the longitude of the ascending node, argument of periapsis, and mean anomaly---are chosen uniformly from $0-2\pi$.  We do not impose a minimum angular separation on our planets, this choice was reported to not alter the orbital crossing time for systems with $\Delta > \Delta_c$ in \citet{obertas}.  We do not observe any systems in two- or three-body resonances with librating resonant angles.

Stability time in this study is interchangeable with close-encounter time, t$_c$.  The time is measured from the integration start to the first close encounter between two planets of less than one Hill radius.  We scale the times by the period of the innermost planet, t$_c$/$T_i$  where $T_i$ is 11.55 days.  Ejections and collisions with the central body are not observed in any runs.

The following three sections use different planet masses and different mass variations which are described in their respective sections.  For each investigation, a ``suite'' of at least 1,000 systems is simulated uniformly across a given range of separations ($\Delta$) with a given prescription for the masses of planets in each system.  The innermost planet is placed at 0.1 AU.  The separation and mass of adjacent planets then determines the semi-major axis of the subsequent planets with Eq.~\ref{eq:space}.

\begin{table}
	\centering
	\caption{Initial conditions for the system and each of the four planets across simulations\label{tab:IC}}
	\begin{tabular}{lr} % four columns, alignment for each		\hline
		\hline
		Parameter & Details\\
		\hline
		Central Mass ($M_{\odot}$)& 1.0 $M_{\odot}$ \\
        Planet Multiplicity & 4 \\
		Planet Mass & Reported in Section \\
		Semi-Major Axis (AU) & $a_1=0.1$ \& Eq.(\ref{eq:space}) \\ 
		Eccentricity & Rayleigh, $\sigma=10^{-6}$ \\ 
		Inclination ($^{\circ}$)  &Rayleigh, $\sigma=10^{-6}$ \\ 
		Arg. of Pericenter ($^{\circ}$) & Random uniform 0-360 \\ 
		Long. of Ascend. Node ($^{\circ}$)& Random uniform 0-360 \\ 
		Mean Anomaly ($^{\circ}$)& Random uniform 0-360 \\ 
		\hline
	\end{tabular}
\end{table}

\section{Stability Time with Spacing Relationship}\label{sec:stability}

\begin{figure}
    \includegraphics[width=\columnwidth]{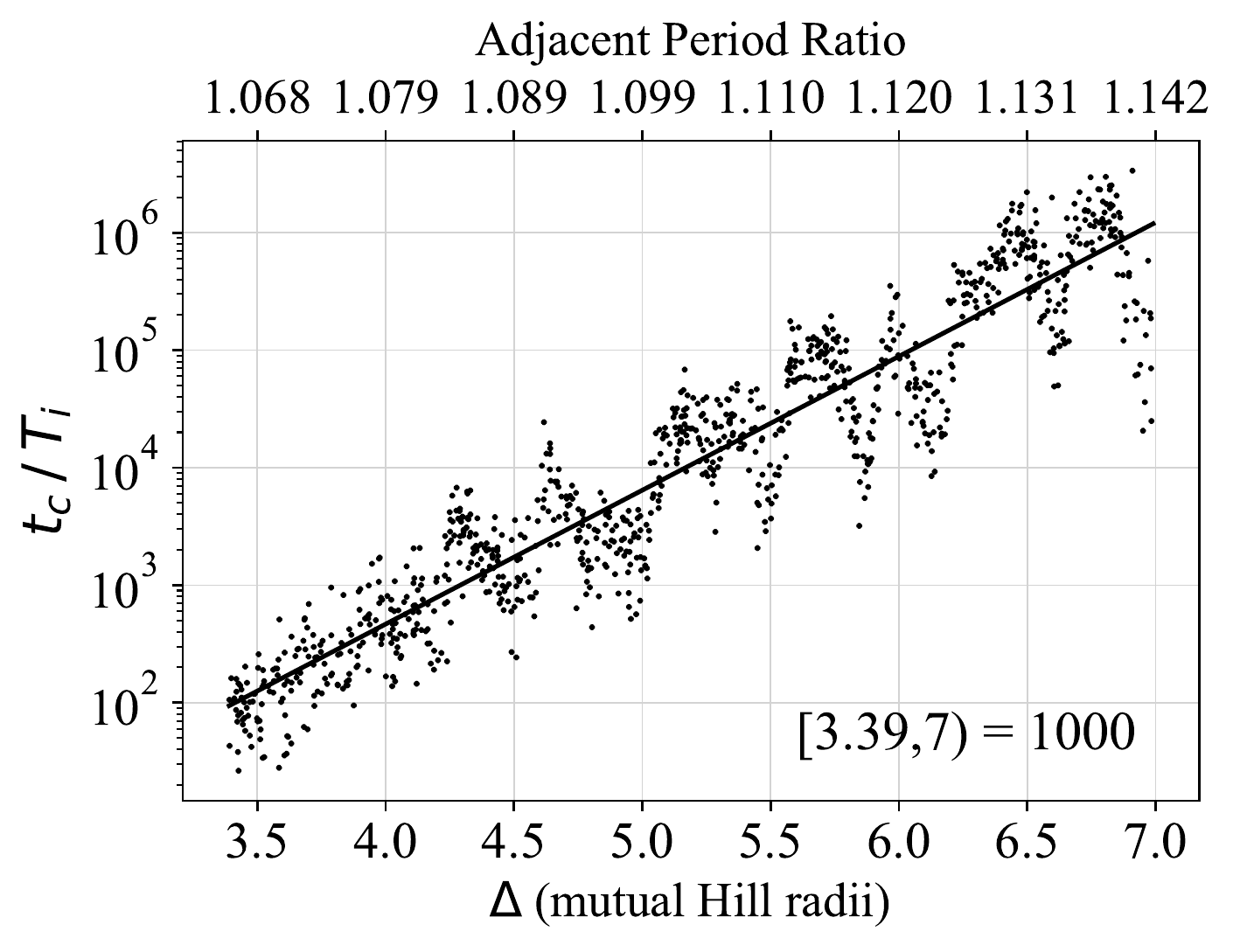}
    \caption{The time to the first close encounter of less than one Hill radius between two planets in terms of the innermost planet's orbital period against spacing measured in mutual Hill radii.  Close encounter (stability) time measured for 1,000 systems of four, equal Earth-mass planets.  Least-squares regression shown between $\Delta_{c}\leq\Delta\leq7$ (Eq.~\ref{eq:delc}).  Top axis shows the period ratio between adjacent planets in the systems.  Note that this axis is not linear and is determined by Eq.~\ref{eq:pratio}. \label{fig:chambers}}
\end{figure}

As an initial test, we show the stability time with spacing relationship for our systems in Fig.~\ref{fig:chambers}.  We run 1,000 simulations of systems with four equal-$1 \mathrm{M_\oplus}$ planets around a Sun-like star with the innermost planet at 0.1 AU.  The separations are chosen uniformly between $\Delta_c$ and 7 R$_\text{H}$ with $\Delta_c\approx3.39$ (Eq.~\ref{eq:delc}).

All simulations had a close encounter within $10^{7}$ orbits of the innermost planet.  The exponential relationship is $log(t_c/T_i)=1.14\Delta-1.89$.  When the maximum separation cutoff is varied between 6.0 and 7.0 mutual Hill radii, the slope ranges from 1.09 to 1.17 and the intercept from -2.04 to 1.69.  These values are similar to the relationship for five planets in \citet{obertas}. 

The top axis in Fig.~\ref{fig:chambers} shows the period ratio of adjacent planets in the systems.  The modulations in the relationship from MMRs are apparent in our data.  For example, the dip around the 1.110 period ratio and 5.57 mutual Hill radii corresponds with a first order MMR with period ratio of 10:9 between adjacent planets in the system.  Not shown in Fig.~\ref{fig:chambers} is the period ratio between next-nearest planets which were also shown to correlate with dips in \citet{obertas}.  There is often a degeneracy in deciding which period ratio is responsible for a dip.

\begin{figure}
    \includegraphics[width=\columnwidth]{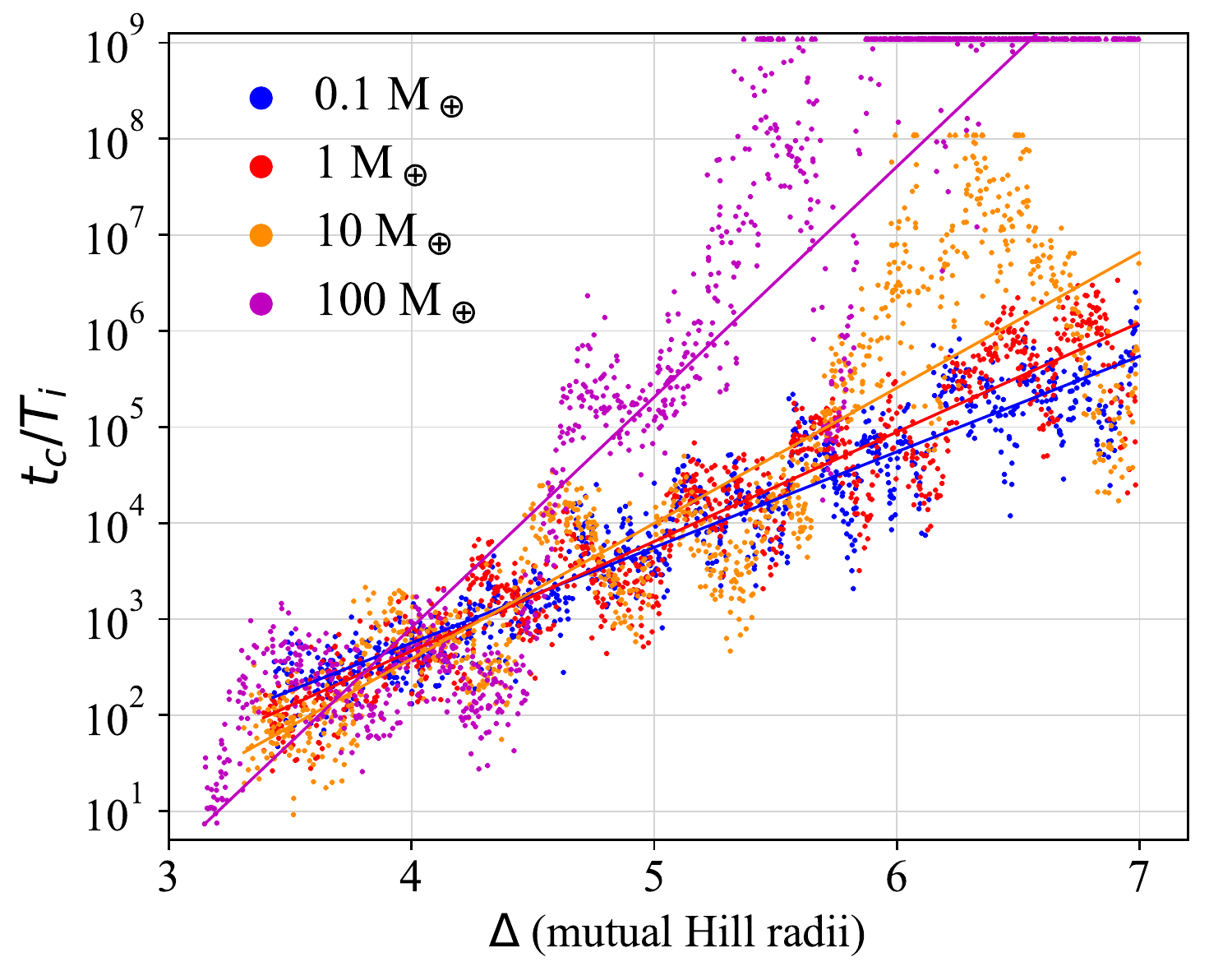}
    \includegraphics[width=\columnwidth]{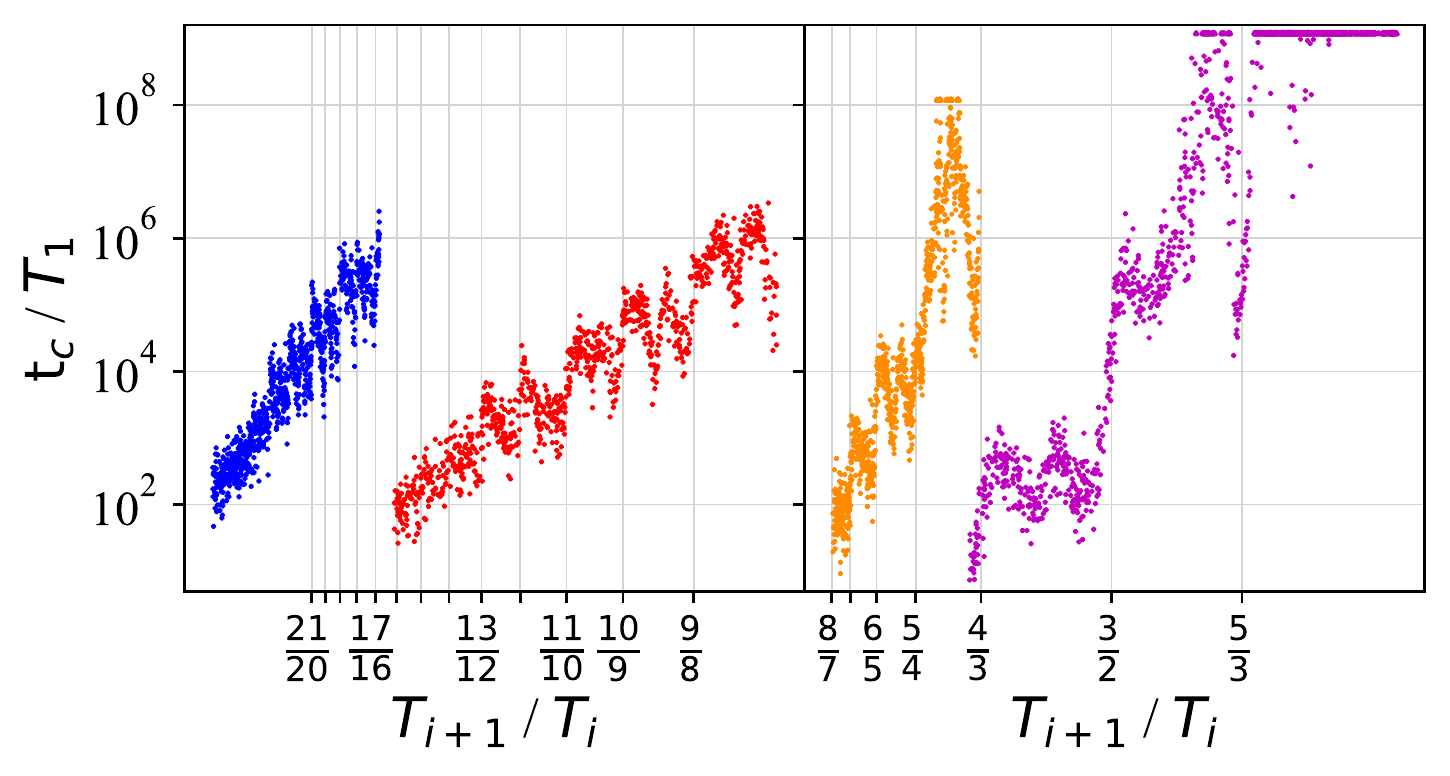}
    \caption{\textit{Top}, stability time of four suites with 1,000 systems of equal-mass planets: 0.1 $\mathrm{M_\oplus}$ (blue), 1 $\mathrm{M_\oplus}$ (red), 10 $\mathrm{M_\oplus}$ (orange), and 100 $\mathrm{M_\oplus}$ (purple) with spacing. 
    Least-squares regressions shown in corresponding color for each suite. 
    \textit{Bottom}, stability time with adjacent planet period ratio.  Systems which reached the maximum integration time are plotted at the maximum integration time of the suite.  Spacings are chosen uniformly with $\Delta_{c}<\Delta<7$.  \label{fig:reschamb}}
\end{figure}

In Fig.~\ref{fig:reschamb}, we show stability times for suites of simulations with equal-$0.1\,\mathrm{M_\oplus}$, equal-$10\,\mathrm{M_\oplus}$, and equal-$100\,\mathrm{M_\oplus}$ planets.  The period ratio of adjacent planets at a given separation is different in each of these suites and results in different modulations of the stability time relationship.  The slope of the stability time relationship increases with mass.  However, the increase in slope is influenced by large modulations.  The 10 M$_\oplus$ suite has a region of systems with longer stability times, a few of which did not have a crossing within the integration time and are plotted at $10^8$ orbits.  The large increase in stability time for the 10 $\mathrm{M_\oplus}$ systems is between the 5:4 ($\Delta\approx5.5$) and 4:3 period ratio ($\Delta\approx7.0$) with a small dip in the middle corresponding to systems with periods near a second order 9:7 period ratio ($\Delta\approx6.2$).  When the maximum separation cutoff is varied between 5.5 and 7.0, the slope ranges from 1.06 to 1.64.  Similarly, the 100 M$_\oplus$ suite has a sharp increase in stability time near the 3:2 ($\Delta\approx4.6$) period ratio.  At smaller separations the stability times of higher mass systems are consistent with lower mass systems.

\section{Variable mass with equal spacing}\label{sec:mass}

We now turn to detailing the stability time relationship for systems with varied planetary mass while keeping equal dynamical spacing in terms of mutual Hill radii as first explored in \citet{chambers}.  Having non-equal mass means that the planets no longer create chains of equal period ratios (See Eq.~\ref{eq:pratio}).  

%This prescription is complimentary to, but not equal to, systems with non-equal separations which is explored in \citet{wu}. 

\begin{figure}
\includegraphics[width=\columnwidth]{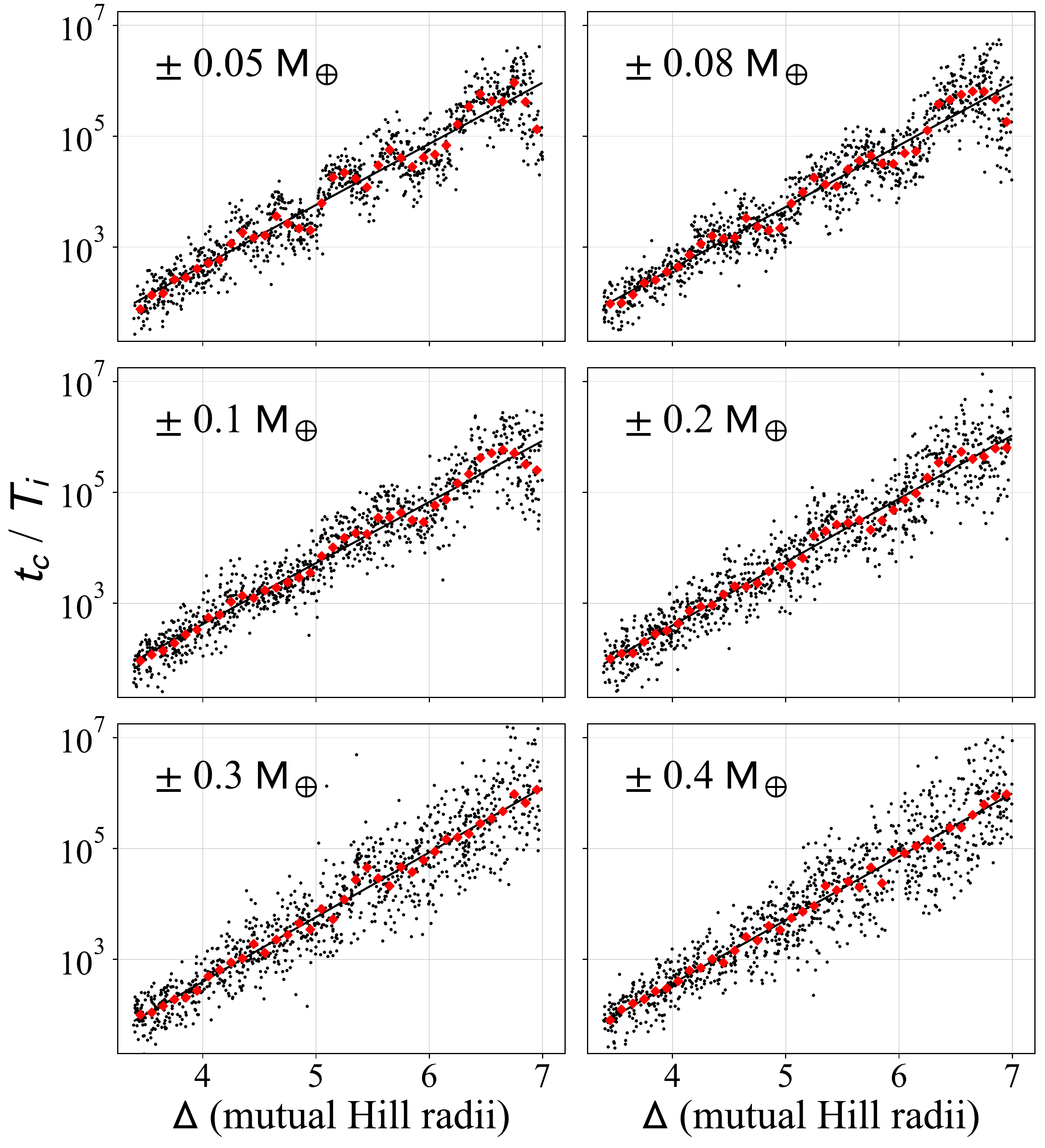}
\caption{Stability time for suites of simulations with non-equal-mass planets.  The mass of each planet in the system is chosen from a normal distribution with $\mu= 1$ $\mathrm{M_\oplus}$ and $\sigma$ shown in the legend.  Systems are spaced by $\Delta$ and by the mutual Hill radius of each set of adjacent planets in the system.  The mean value of stability time for systems in bins 0.1$\Delta$ wide are shown in red.  Compare to the equal-mass systems of Fig \ref{fig:chambers}. \label{fig:varmass}}
\end{figure}

The mass of each planet in four-planet systems is chosen from a normal distribution, $\mathcal{N}(\mu,\sigma^2)$.  The mean value is kept at one Earth-mass while we vary the standard deviation across the suites up to 50 per cent of the mean value.  With a small multiplicity, the deviation in mass of individual systems will have deviations both smaller and larger than the distribution from which they were chosen. We require each system to have non-negative masses and a mean and standard deviation within 5 per cent of the target value.  Systems are simulated randomly with $\Delta_{c, 1\mathrm{M_\oplus}}\le\Delta\le7$.  The innermost planet is at 0.1 AU.  The semi-major axes of subsequent planets are given by the chosen $\Delta$, the mass of the planet, and the mass of it's interior neighbor (Eq.~\ref{eq:space}).  All simulations had a close encounter within the max integration time of $10^{8}$ orbits of the innermost planet.  

The results are shown in Fig.~\ref{fig:varmass}.  \citet{chambers} finds that non-equal masses decreases stability time at large separations.  In our mass range, the slope of the stability relationship is only decreased by 3.5 per cent and the y-intercept increased by 6.3 per cent in the $\sigma=0.3$ $\mathrm{M_\oplus}$ suite compared to the equal-mass suite.  Lower mass planets in non-uniform mass systems gain eccentricity quicker than planets in uniform mass systems leading to this slight decrease in stability time \citep{chambers}.

\begin{figure}
\includegraphics[width=\columnwidth]{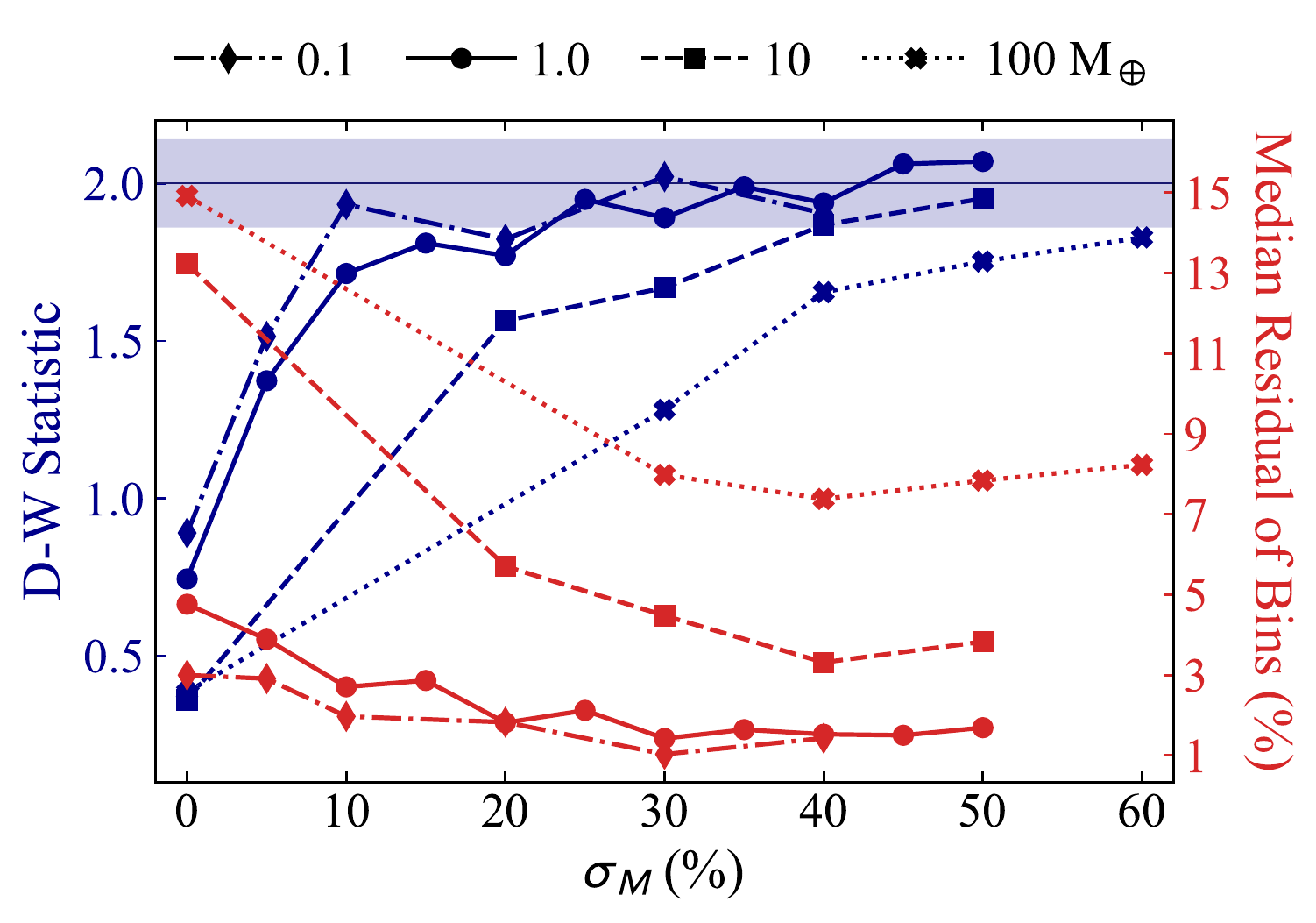}
\caption{Two measurements of the structure in the stability time relationship---Durbin-Watson test statistic (\textit{blue, left)} and the median residual of bins (\textit{red, right})---for suites with varying mean mass (\textit{linestyle)} and standard deviation in mass (\textit{x-axis}).  \textit{Shaded blue} region indicates the values of the D-W test where there is no significant autocorrelation.  The median residual of bins measures the average absolute vertical distance of the red dots (binned stability time) to the line (expected stability time) in the corresponding suite in Fig.~\ref{fig:varmass} as a percentage of the expected stability time.
}\label{fig:structure}
\end{figure}

As the standard deviation of the mass distribution increases, the structure in the stability time relationship is erased.  We measure the decrease in structure in two ways shown in Fig.~\ref{fig:structure}.  A Durbin-Watson (D-W) test on residuals returns a test statistic that ranges from 0-4 with a value of 2.0$\pm$0.14 indicating no autocorrelation at the one per cent significant level for a sample size of 1000 \citep{DWTest}.  We perform this test on the log-residuals of stability time, \(\log\frac{t_{c}}{T_i}-\log\hat{\frac{t_c}{T_i}}\) where $\hat{\frac{t_c}{T_i}}$ is the expected time from the regression.  The test statistic for the uniform 1 M$_\oplus$ mass suite is 0.74 indicating positive autocorrelation.  When the standard deviation is above 0.25 M$_\oplus$ the D-W test indicates no autocorrelation.

The second measure of structure is the residual of binned stability times.  We collect systems in bins with widths of 0.1$\Delta$ and measure the mean stability time in each bin (\textit{red points} in Fig.~\ref{fig:varmass}).  We measure the absolute residual of these binned times, \(\vert\log\frac{ \overline{t_{c}}_\textrm{,bin}}{T_i}-\log\hat{\frac{t_c}{T_i}}\vert\), as a percent of the expected stability time and report the median across all bins.  The binned residuals decrease with increasing variation, reaching a minimum in the $\sigma=0.3$ $\mathrm{M_\oplus}$ suite.  This variation has a median binned residual (1.42 per cent) that is 3 times smaller than that of the equal-mass systems (4.76 per cent). 

We measure the dispersion of stability time in the same bins as above.  The mean standard deviation in the uniform mass suite is 0.28 dex and shows no correlation with spacing.  In systems with 1$\pm$0.3 M$_\oplus$ the stability time distribution at given spacing is much wider with bins having an average deviation of 0.44 dex.  The deviation shows an increase with spacing from 0.22 dex at $\Delta_c$ to 0.64 dex at $\Delta=7$.  The trend is $\sigma=0.11\Delta-0.12$ with a correlation coefficient of 0.84.

We follow the above procedure with systems with mean mass of 0.1 M$_\oplus$, 10 M$_\oplus$, and 100 M$_\oplus$ and varying standard deviation.  For each mass and deviation 1000 systems are generated with $\Delta_{c, \mathrm{\overline{M}}}\le\Delta\le7$.  The suites with mean mass of 100 M$_\oplus$ have a maximum separation of $\Delta=6$ and 4\% of systems have stability times longer than the maximum integration time of 10$^9$ orbits and are removed from the analysis.

Increasing the variation in mass decreases the structure for all planet mass.  The amount of variation needed to return no autocorrelation and minimize the binned residuals is $\sigma_M=0.1-0.2~\mathrm{\overline{M}}$ for the average 0.1 M$_\oplus$ systems.  The variation increases by approximately 10 per cent for a tenfold increase in mass.  For the 100 M$_\oplus$ systems, some structure still remains for a $\pm$60 M$_\oplus$ variation.  The increase in stability time between the 8:5 ($\Delta\approx5.3$) and 5:3 ($\Delta\approx5.8$) period ratio is over 2 orders of magnitude larger than the predicted time from a linear relationship.  The mass variation needed to erase the modulations increases with lower integer resonances (period ratios closer to 2:1).  This is discussed further in Sect.~\ref{sec:pratio}.

\begin{figure}
\includegraphics[width=\columnwidth]{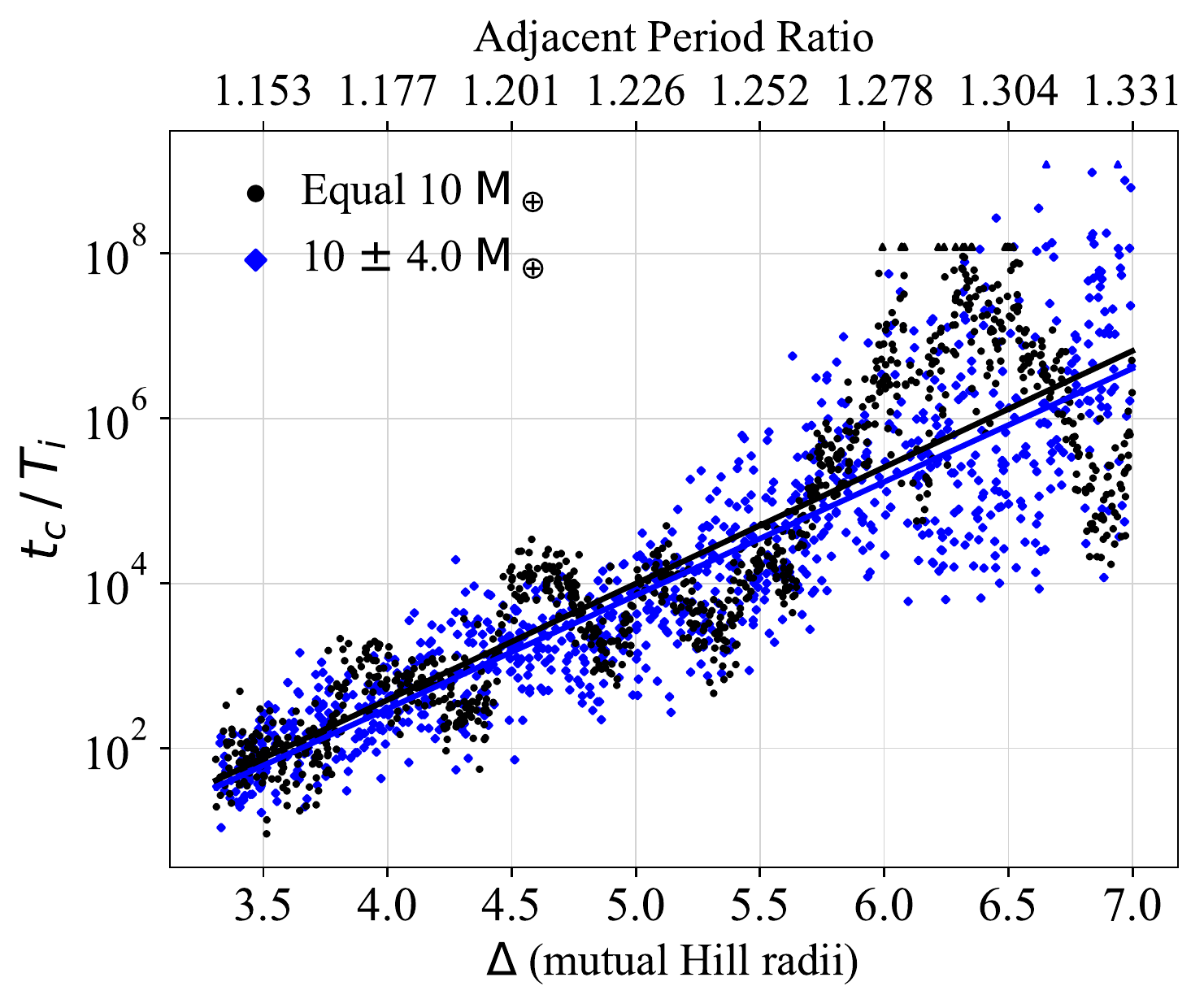}
\caption{Stability time for systems with equal-10 M$_\oplus$ planets (\textit{black}) and for systems with masses drawn from  $\mathcal{N}(10\,\mathrm{M_\oplus},(4\,\mathrm{M_\oplus})^2)$ (\textit{blue}) with least-squares regression.  Maximum integration time is 10$^8$ orbits for equal and 10$^9$ orbits for non-equal.  Top axis shows the period ratio between adjacent planets in the equal-mass systems.\label{fig:searth}}
\end{figure}

\subsection{Planet Multiplicity}\label{sec:multi}

In \citet{chambers}, the slope of the stability time relationship decreases with planet multiplicity up to five-planet systems and then stays relatively constant up to 20-planet systems.  They note that the modulations from period ratio commensurability become more apparent with increasing planet multiplicity.  However, their low resolution, with under 105 systems per suite, limits their ability to analyze the modulations.

\begin{figure}
\includegraphics[width=\columnwidth]{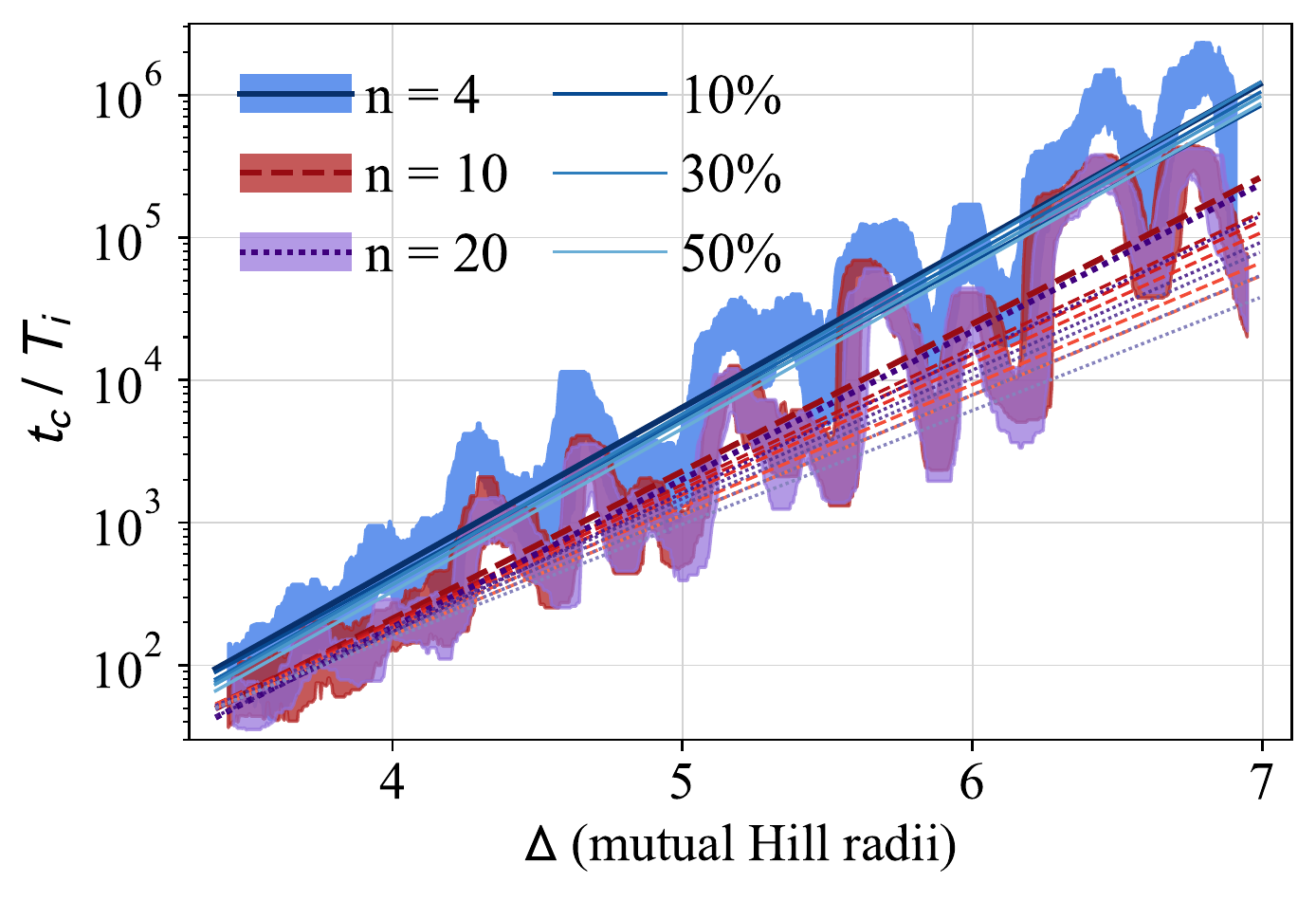}
\includegraphics[width=\columnwidth]{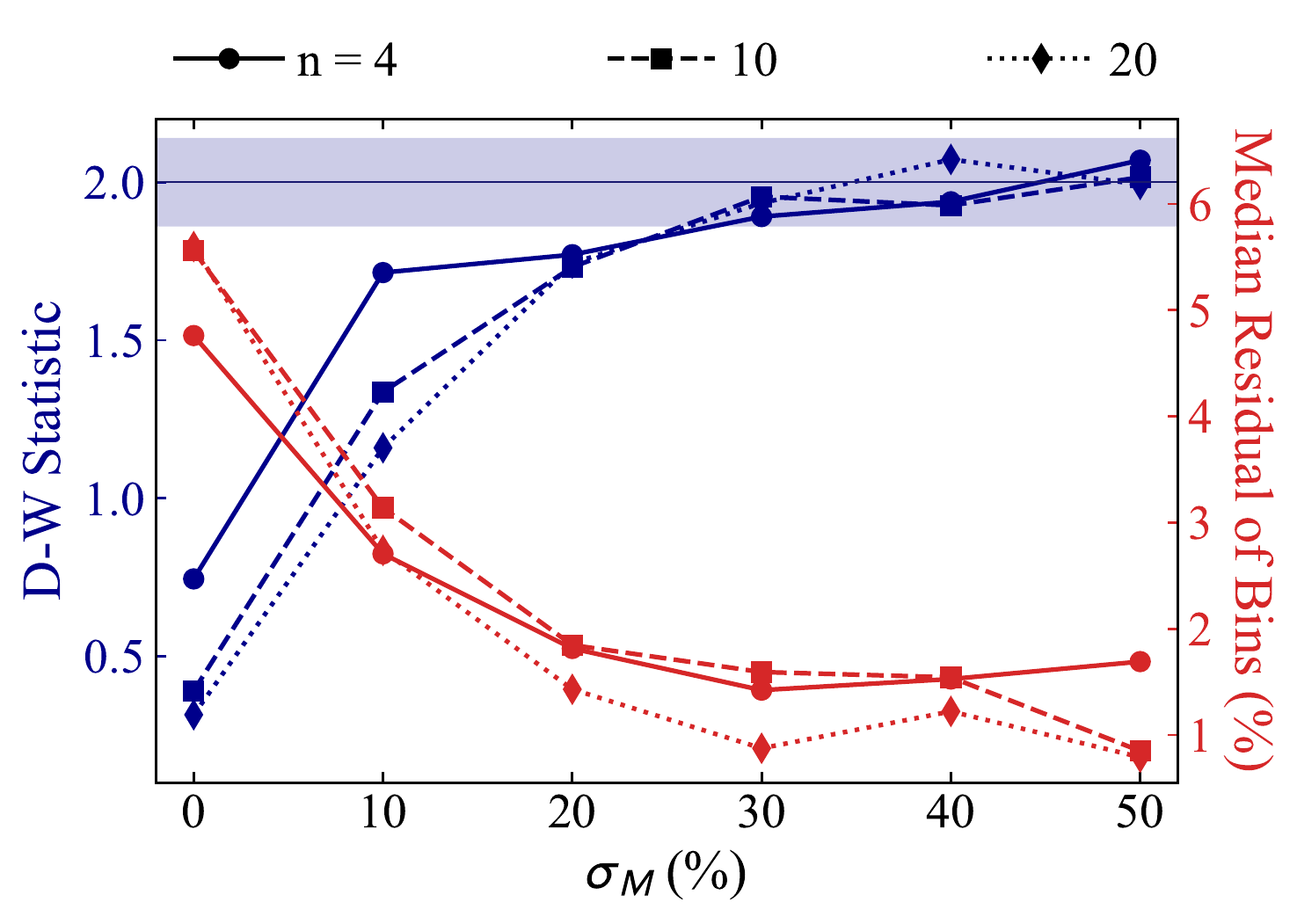}
\caption{\textit{Top}, stability time with spacing for three suites of varying planet multiplicity.  The shaded region represents a running $\pm1\sigma$ region of stability time.  Least-square regression shown in corresponding color for each suite.  Lines with successively lighter-shades of each color represent linear regressions of varied mass suites with deviations of 0.1, 0.2, 0.3, 0.4, and 0.5 M$_\oplus$.  \textit{Bottom}, two measurements of the structure in the stability time relationship---Durbin-Watson test statistic (\textit{blue, left)} and the median residual of bins (\textit{red, right})---for suites with mean mass of 1 $M_\oplus$ and varying planet multiplicity (\textit{linestyle)} and standard deviation in mass (\textit{x-axis}).  \textit{Shaded blue} region indicates the values of the D-W test where there is no significant autocorrelation.}\label{fig:multistruc}
\end{figure}

We compare our 4 planet systems with 10-, and 20-planet systems with planets of 1 M$_\oplus$.  Shown in Fig.~\ref{fig:multistruc} \textit{top}, the slope decreases with more planets while the intercept remains consistent.  The exponential relationship for 10 planets is $log(t_c/T_i)=1.03\Delta-1.81$ and 20 planets is $log(t_c/T_i)=1.04\Delta-1.89$.  The modulations do not get larger, as in the residuals do not significantly increase, but the spread at a given spacing gets smaller which makes the modulations more apparent.  The average standard deviation in 0.1$\Delta$ wide bins for 4 planet systems is 0.28 dex, 10 planets is 0.25 dex, and 20 planets is 0.23 dex.

The width is in agreement with the 0.22 dex standard deviation of the stability time distribution for systems of four Neptune-mass planets separated by 5 mutual Hill radii from \citet{Rice18}.  \citet{Hussain2020} shows that the 0.22 dex deviation matches the 5-planet systems of \citet{obertas}.  However, \citet{Hussain2020} also finds that the standard deviation of ``shadow integrations'' (slight changes on initial conditions of a single system across a wide range of spacings, masses, eccentricities, and inclinations) has a preferred value of 0.43$\pm$0.16 dex for three planet systems.   We find the smaller 0.22 dex deviation better-matches that of \citet{obertas} because of the higher planet multiplicity of \citet{Rice18} and \citet{obertas}.  For systems with three Earth-mass planets between $\Delta_{c}\le\Delta\le7$ we find the deviation in 0.1 $\Delta$ wide bins to be 0.36 dex.

We simulate varied mass suites where each system's 4-, 10- or 20-planet masses are chosen from a normal distribution centered at 1 M$_\oplus$ with a 10-50 per cent standard deviation.  We require each system to have a mean and standard deviation withing 5 per cent of the target value.  The D-W test shows no significant correlation and the median residual of bins is below 2 per cent with a 30 per cent variation (0.3 M$_\oplus$) for all planet multiplicities.

\subsection{Change in Period Ratio}\label{sec:pratio}

In reality, a planetary system will have non-uniform masses, dynamical spacings, and period ratios.  In Eq.~\ref{eq:pratio}, we see all these quantities are related.  Thus a variation in the planets' masses correspond to a variation in period ratio.  If we propagate the previously found standard deviations of mass in the 4-planet systems we can find the deviation in period ratios which disrupts the chain of MMRs resulting in wide distributions of stability time.  This deviation will also depend on the separation, $\Delta$. 

We use the variance formula to describe the period ratio deviation caused my a mass variation,
\begin{equation}\label{eq:pvar} 
\sigma_{P_{r}}=\sqrt{\left(\frac{\delta P_{r}}{\delta K}\right)^2 \sigma_K^2},
\end{equation}
where $P_{r}$ is given in Eq.~\ref{eq:pratio}.  The variance of K, $\sigma^2_K$, is given by the derivative of Eq.~\ref{eq:K} with respect to $m_1+m_2$ multiplied by $\sigma_{m}$.  We also include a term under the square root in Eq.~\ref{eq:pvar} for the deviation in $\Delta$ from our binning process, but this term is negligible for $K\ll\Delta$. 

\begin{figure}
\includegraphics[width=\columnwidth]{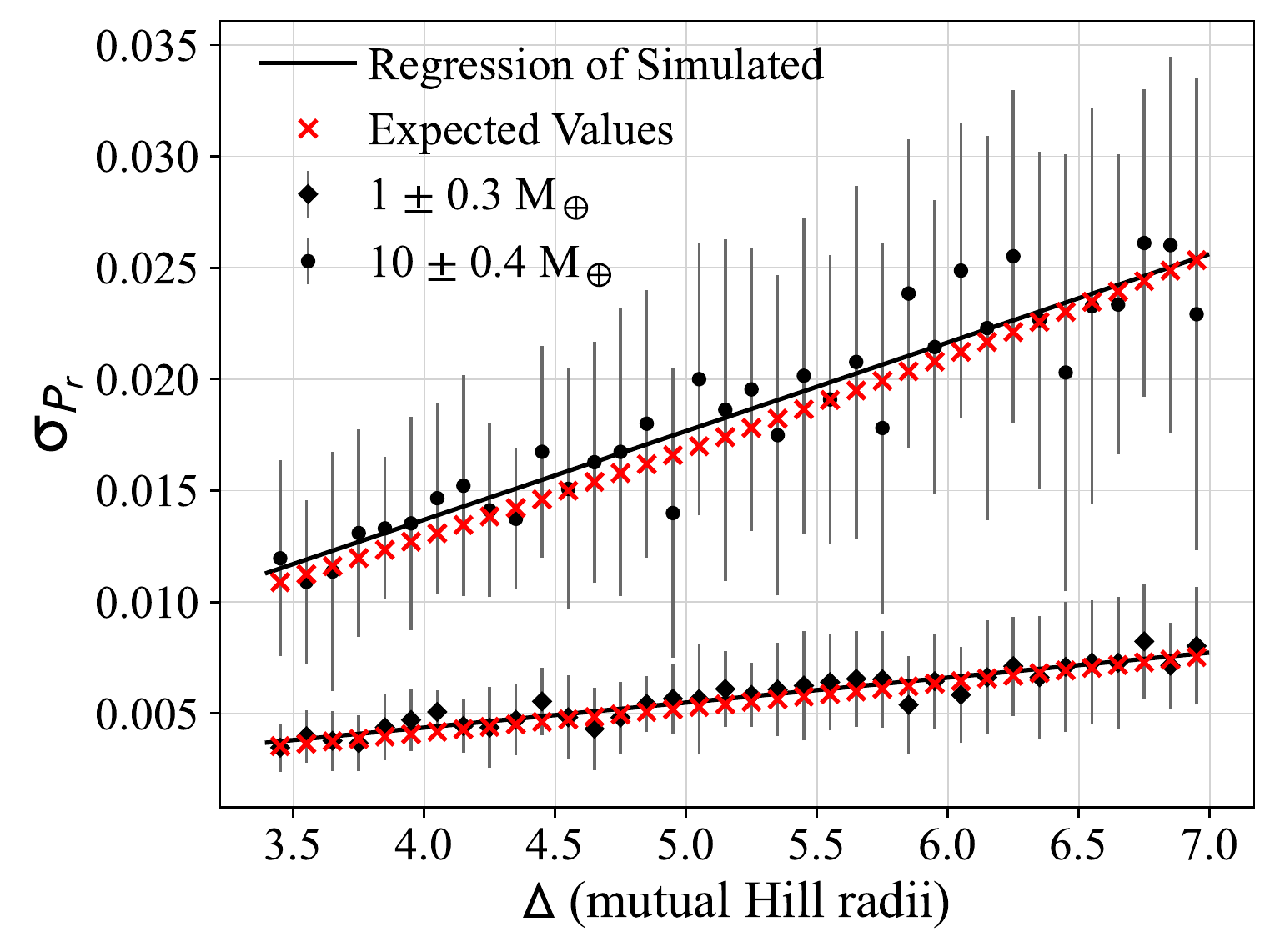}
\caption{The period ratio are measured between every pair of adjacent planets in the 4-planet systems with masses of 1$\pm$0.3 M$_\oplus$ and 10$\pm$4 M$_\oplus$.  Shown here is the average standard deviation of period ratio and 1-$\sigma$ uncertainties for systems in 0.1 wide bins of $\Delta$.  Black line shows the regression of the simulated standard deviations while red crosses show the expected deviation from error propagation using the variance formula. \label{fig:periodsig}}
\end{figure}

For systems with masses drawn from $\mathcal{N}(1\,\mathrm{M_\oplus},(0.3\,\mathrm{M_\oplus})^2)$ and $\mathcal{N}(10\,\mathrm{M_\oplus},(4\,\mathrm{M_\oplus})^2)$, the expected deviations in the period ratios from Eq.~\ref{eq:pvar} are plotted on Fig.~\ref{fig:periodsig}.  The deviation is fairly-linear with separation in this region.  We measure the three period ratios in each 4-planet system in both suites.  The standard deviation of period ratios is then measured for each systems and averaged in bins with width of 0.1$\Delta$.  There are on average 28 systems in a bin.  As shown, a linear regression of the deviations are consistent with the expected values from Eq.~\ref{eq:pvar}.  

To illustrate this result's affect on stability time, consider the dip at $\Delta=5.5$ in Fig.~\ref{fig:chambers} which we claim is caused by being near a 10:9 MMR, $P_{r}\approx$ 1.11.  The average system with this separation would likely have a stability time which is 10,000 orbits less than expected from the trend line (13.4 versus 23.9 thousand orbits).  A normal deviation in the period ratio of the planets of $\sim$0.006 (0.5 per cent) is enough for the average system to remain stable for longer---removing the dip and returning the average system to the trend line.  Predictions of stability time for real systems with these small variations are best described by the exponential relationship and not the modulations.

This deviation in period ratio caused by a mass variation in an equal-spacing system is spacing and mass dependent.  Thus systems with planets near period ratios with smaller integer ratios (closer to 2:1) have larger period ratio deviations.  A system with masses of 10$\pm$3 M$_\oplus$ would have a period deviation of 0.014 which is over twice that of a 1$\pm$0.3 M$_\oplus$ system.  However, we find that even larger mass deviations and thus large period deviations are needed to erase modulations caused by lower integer period ratios.  At $\Delta=5.5$ planets of 10 M$_\oplus$ are near a 5:4 period ratio.  We find that a $\pm$4 M$_\oplus$ deviation erases the modulation.  As shown in Fig.~\ref{fig:periodsig} this corresponds to a 0.019 (1.5 per cent) period ratio deviation.

Inhomogenous systems with variations in period ratios as large as the ones detailed here have wide distribution of stability times for a given spacing. Understand this difference in stability times in non-equal mass systems compared to equal mass systems requires an understanding of the mechanisms driving instability.  Proposed mechanisms which drive the instability resulting in the stability time with spacing relationship include those by \citet{zhou} and \citet{yalinewich}.  Of promise is the theory proposed by \citet{quillen} and \citet{Petit2020}; they propose that overlapping three-body resonances drives the chaos.  \citet{Petit2020} show that this mechanism predicts an exponential relationship at small separations and a rapid increase in stability time at larger separations (this increase is discussed in Sect.~\ref{sec:long}).  Shorter stability times caused by nearness to two-body resonance is then seen as a dip away from the predicted stability from three-body resonances.  In varied-mass system, with unequal period ratios a more diverse set of two-body resonances are possible which results in this wider distribution of stability times.  

\begin{figure}
\includegraphics[width=\columnwidth]{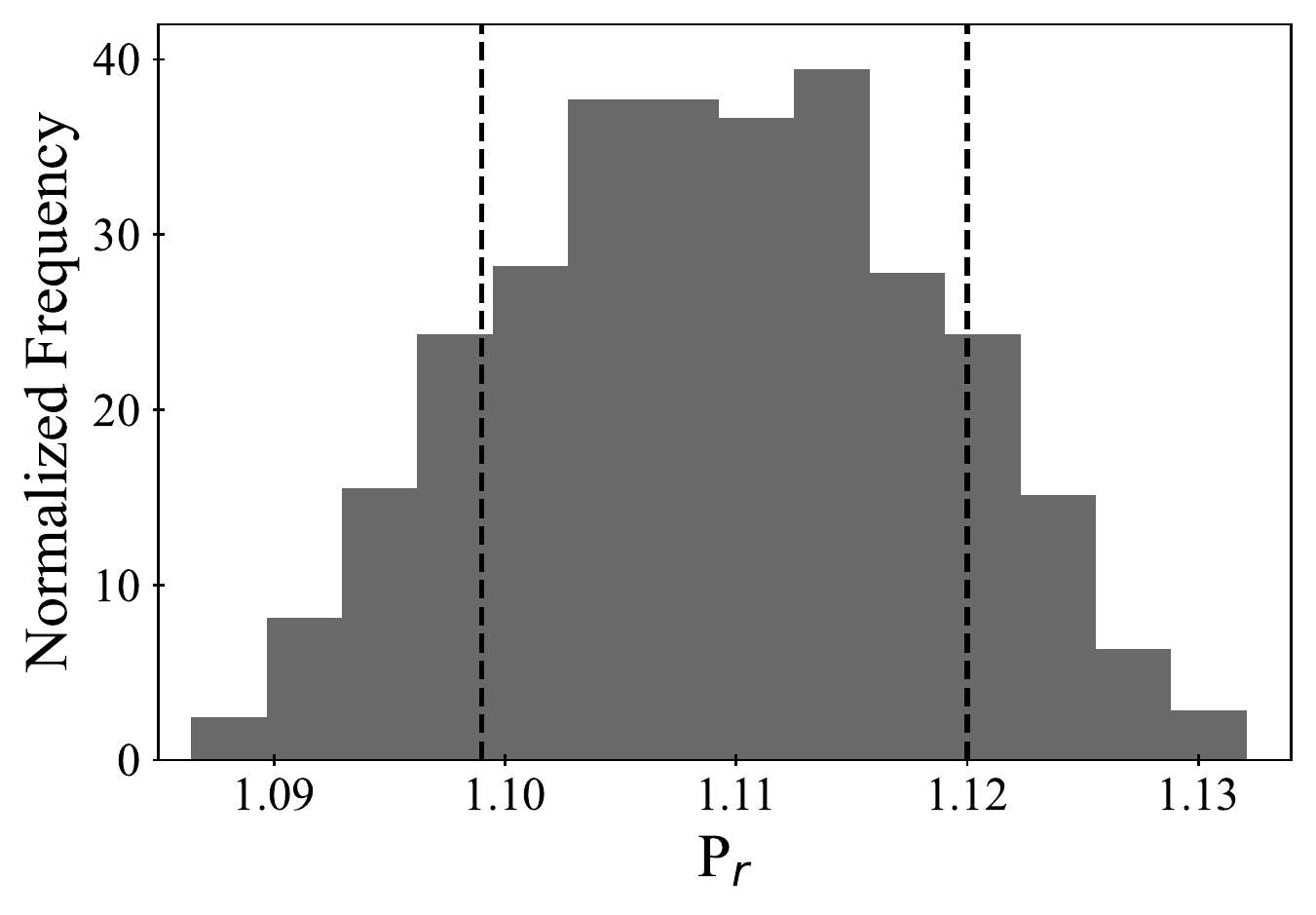}
\caption{Frequency of period ratios between every pair of adjacent planets in the 4-planet systems with masses of 1$\pm$0.3 M$_\oplus$ with uniform separations of $5<\Delta<6$.  Vertical dashed lines show the period ratio of $\Delta=5$ and $\Delta=6$ for uniform mass systems. \label{fig:perioddist}}
\end{figure}

For our simulations, Fig.~\ref{fig:perioddist} shows the frequency of period ratios in the 290 systems with separations of $5<\Delta<6$ with masses of 1$\pm$0.3 M$_\oplus$.  In the uniform mass systems these ratios are uniform between 1.099 and and 1.12 (see Fig.~\ref{fig:chambers}).  Period ratios that were only available for $\Delta=4.5$ or $\Delta=6.5$ systems are now available to drive the instability for the varied mass systems.

\section{Systems with masses from 1-10 Earth-masses}\label{sec:hetero}

\begin{figure}
\begin{center}
\includegraphics[width=\columnwidth]{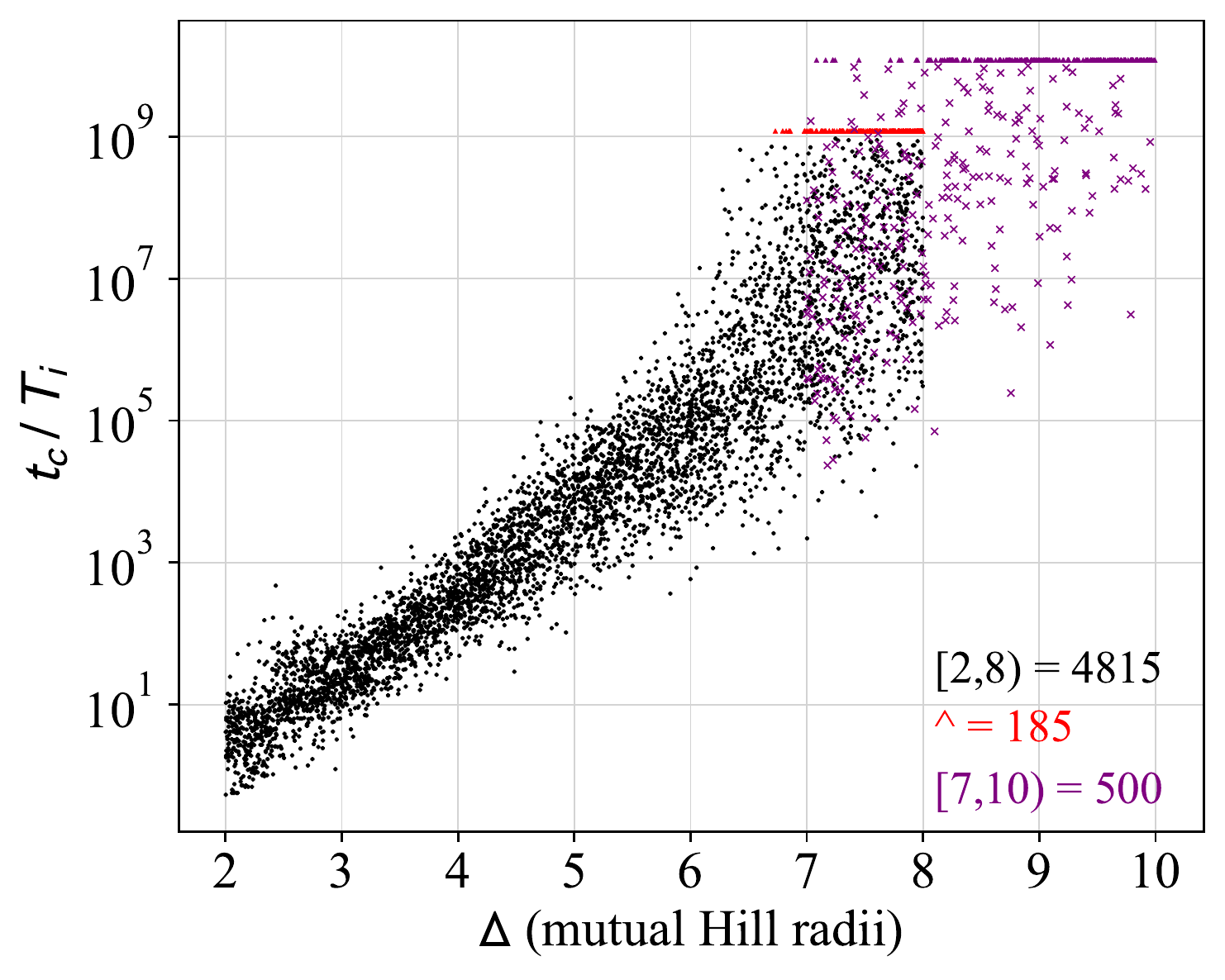}
\end{center}
\caption{Stability time in terms of the innermost planet's period against the spacing of the system in mutual hill radii of adjacent planets.  Each planet's mass in the four planet systems is chosen uniformly from 1 M$_\oplus$ to 10 M$_\oplus$.  5,000 systems (\textit{black dots}) are simulated with $\Delta=[2,8)$ and maximum integration time of $10^9$ orbits.  An additional 500 systems (\textit{purple x's}) are simulated with $\Delta=[7,10)$ and maximum integration time of $10^{10}$ orbits.  Systems shown as triangles did not have a close-encounter of less than 1 Hill Radius within the maximum integration time and are plotted at the maximum.}
\label{fig:fullchamb}
\end{figure}

As seen in Section \ref{sec:mass}, adding a normal variation to the individual planetary mass in systems with equal mutual Hill radii spacing smooths over the modulations from systems with period ratios near MMR.   We take our simulations further by varying mass by up to a factor of ten.  Each planet's mass in the four planet system is chosen from a uniform distribution between 1 and 10 $\mathrm{M_\oplus}$.  The stability time for 5,000 of these heterogeneous-mass systems are shown in Fig.~\ref{fig:fullchamb} with a maximum integration time of $10^{9}$ orbits.  96.3 per cent of systems remain stable throughout the entire integration.  The standard deviation of mass in these systems is 40$\pm$17 per cent of each system's mean mass; larger in the majority of systems than the deviations which erase modulations in this mass range in Section \ref{sec:mass}.  The exponential relationship between the spacing and the stability time is maintained for these systems, and no MMR structures are visually apparent.  In addition, we simulate 500 systems that extend our spacing to $\Delta=10$ and integration time to $10^{10}$ orbits.

\begin{figure}
\begin{center}
\includegraphics[width=\columnwidth]{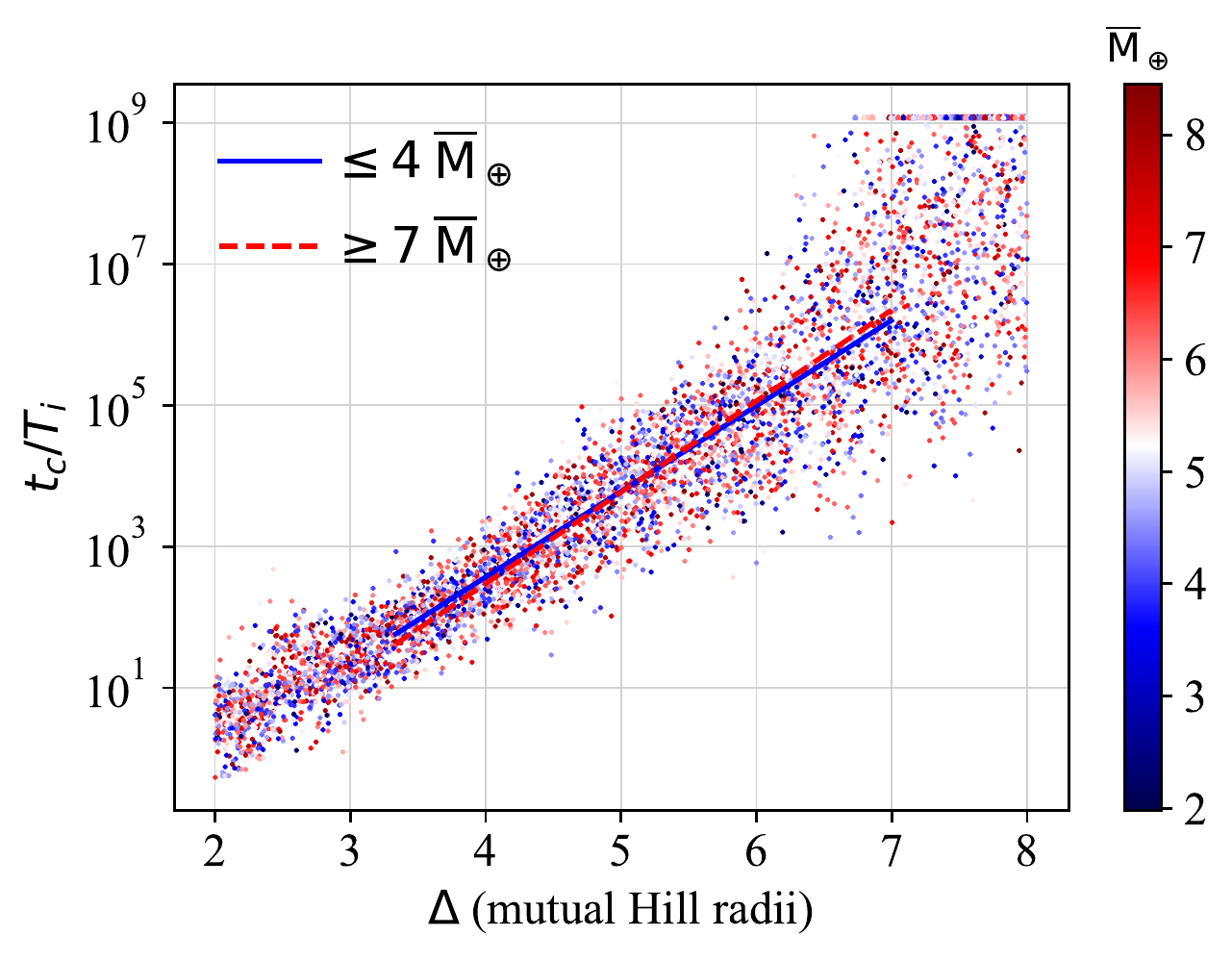}
\includegraphics[width=\columnwidth]{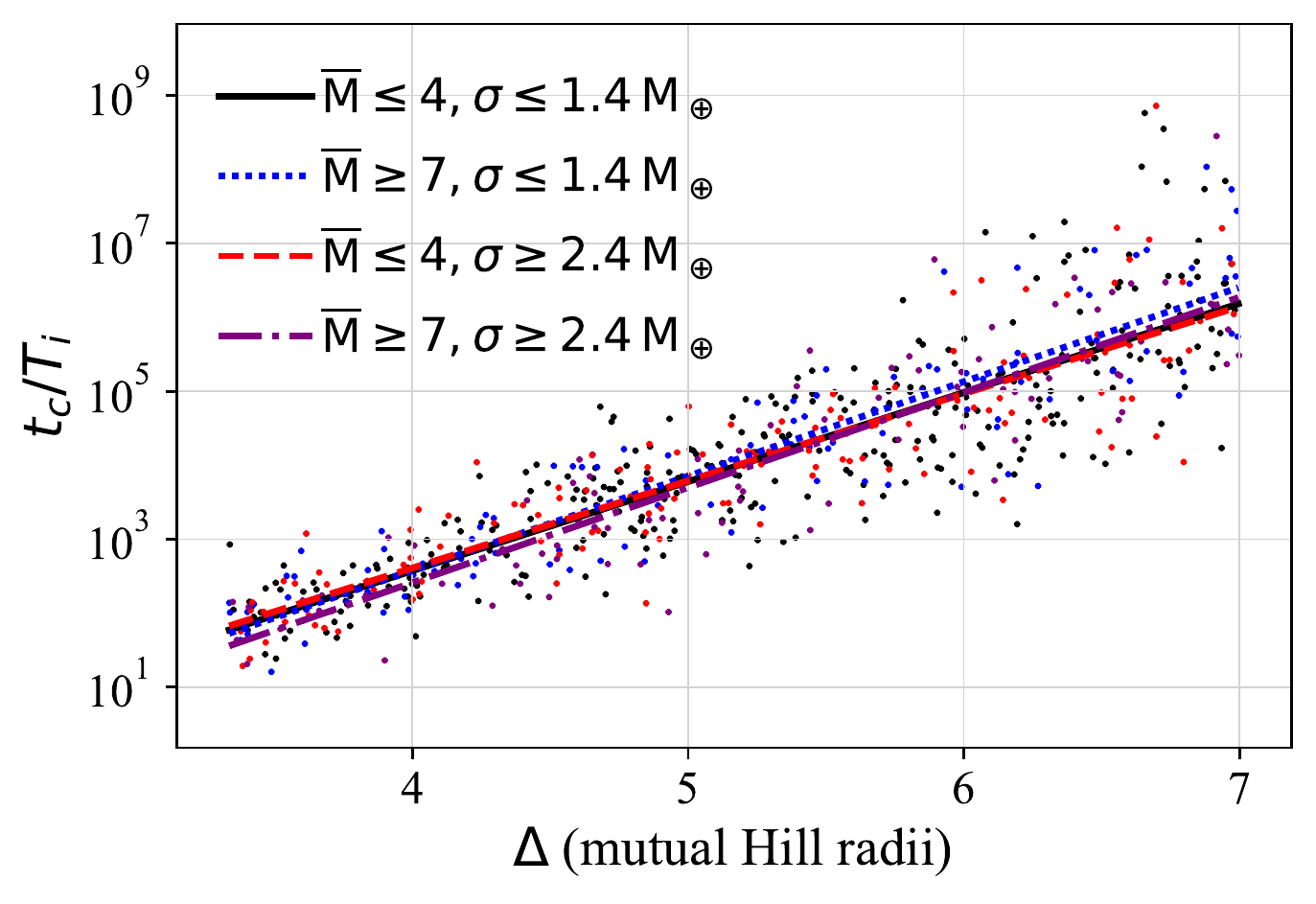}
\end{center}
\caption{\textit{Top}, the systems from Fig.~\ref{fig:fullchamb} are colored by the average mass of their 4 planets, and linear regressions are shown for systems with average mass $\geq7$ $\mathrm{M_\oplus}$ and $\leq4$ $\mathrm{M_\oplus}$ between $\Delta_{c}<\Delta<7$. \textit{Bottom}, systems with average mass $\geq7$ $\mathrm{M_\oplus}$ and $\leq4$ $\mathrm{M_\oplus}$ are divided into systems with low standard deviation in mass, $\leq1.4$ $\mathrm{M_\oplus}$, and high standard deviation, $\geq2.4$ $\mathrm{M_\oplus}$.  The colors of the points correspond with the colors of the plotted linear regression. 
 Each group has between 90 and 300 systems.}
\label{fig:mass}
\end{figure}

In Fig.~\ref{fig:mass} \textit{top}, systems are colored by their average mass.  Linear regressions for systems with an average mass above 7 $\mathrm{M_\oplus}$ and below 4 $\mathrm{M_\oplus}$ and with $\Delta_{c, 5.5\mathrm{M_\oplus}}<\Delta<7$ return similar slopes and intercepts.  The per cent difference in the slope of these two regressions is 5 per cent.  This is much smaller than the 20 per cent difference in slopes in the uniform 1 and 10 M$_\oplus$ suites in Section \ref{sec:stability}.  Although \citet{zhou} with 9-planet systems predicts a lower stability time with a power law relationship, the empirical formula in \citet{zhou} predicts times at $\Delta$=7 of 10$^{5.6}$ for 1 M$_\oplus$ and 10$^{5.9}$ for 10 M$_\oplus$.  The difference in their times in log-space is comparable to our high and low mass systems which have average times of  10$^{6.2}$ and 10$^{6.4}$.

Furthermore, in Fig.~\ref{fig:mass} \textit{bottom}, we split systems in roughly upper thirds and lower thirds of average mass and split those groups into upper and lower thirds of standard deviation in mass.  These groups only have 90-300 systems.  All groups return similar slopes and intercepts.  D-W tests on the log-residuals of stability time of the above 7 $\mathrm{M_\oplus}$ and below 4 $\mathrm{M_\oplus}$ systems and the four groups here all return test statistics above 1.8 indicating no autocorrelation.

The primary predictor of stability time remains the spacing between the planets.  It is equally likely that the encounter pair contains and does not contain the minimum mass planet in the 4-planet systems.  As in \citet{Rice18} the inner pair is over 10 per cent more likely to collide than the outer pairs.  A further explanation of the instability pair would likely need to weigh the position of the pair in the planetary system, the pair's masses, the other masses in the systems, and the nearness of each pair to 2- and 3-body resonances.  \citet{Tamayo2020} shows machine learning is a powerful tool in predicting stability times for this multi-faceted problem.

%\textit{Right}, a violin plot showing the distribution of stability times for all systems in 0.2 $\Delta$ wide bins.  Means and medians of the bins are plotted for the systems up to $\Delta=7$.  Larger values of $\Delta$ are under-sampled; the average amount of systems in the bins of corresponding color are shown in the bottom left

\subsection{The Heteroscedasticity of Stability Time}\label{sec:subhetero}

As the period ratios increase, the difference in a near-resonant system's stability time from the expected exponential relationship seems to grow.  Consider in Fig.~\ref{fig:chambers} the modulation from the 10:9 (1.110) versus that of the 8:7 (1.142) period ratio.  The modulations are smoothed over when using varying masses, Fig.~\ref{fig:varmass}, but nonetheless the variance of stability time at a given separation increases with said separation as found in the 1$\pm$0.3 M$_\oplus$ systems.  

\begin{figure}
\begin{center}
\includegraphics[width=\columnwidth]{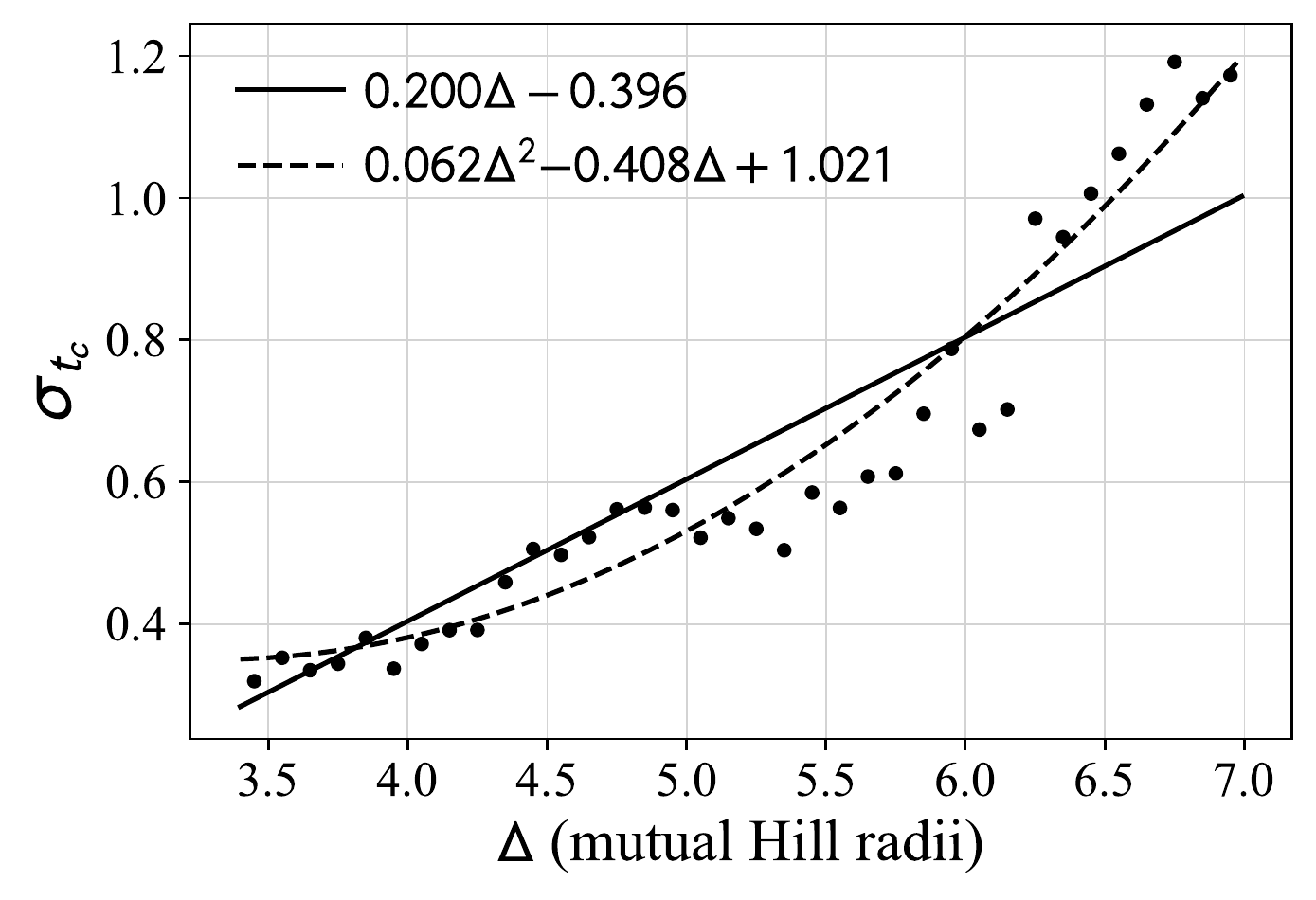}
\end{center}
\caption{The standard deviations of stability time in bins 0.1 $\Delta$ wide are measured for the 3,034 systems with $\Delta_{c}<\Delta<7$. 
 Each bin has over 70 systems.  The measured values from the simulations are compared to the linear and quadratic standard deviation models.  Although it is not clear visually which model is better, the data is clearly heteroscedastic.}
\label{fig:sigma}
\end{figure}

With a range of masses from 1-10 M$_\oplus$, this increase is apparent.  The dispersion of stability time with separation is shown in Fig.~\ref{fig:sigma}.  Systems with the same separation but different masses have stability times distributed with a standard deviation from 0.32 up to 1.20 dex.  These deviations are 50-100 per cent larger than in the 1$\pm$0.3 M$_\oplus$ systems.  Thus the stability times of systems with $\Delta=7$ vary from $10^3-10^9$ orbits of the innermost planet. 

Data with non-constant dispersion is referred to as ``heteroscedastic''.  The linear least-squares regression assumes the residuals are normally distributed and have a constant standard deviation.  We build a Bayesian Linear Regression that maximizes the coefficient of determination, $r^2$, by sampling values for the slope, intercept, and standard deviation using PyMC3 \citet{pymc}.  The slope and intercept are sampled from wide normal distributions centered at 0 with variance of 10$^8$ while the standard deviation is sampled from wide, positive half normal distributions.  To do so we use a Markov Chain Monte Carlo (MCMC) with a number of chains equal to the number of parameters and at least a total of 10,000 sample draws across all chains.  

We analyze the 3,034 systems between $\Delta_{c, 5.5\mathrm{M_\oplus}}\approx3.34$ and $\Delta=7$.  In this region less than 0.3 per cent of systems did not have a close encounter within our integration time and are given a stability time of $10^9$ orbits for analysis.  A Bayesian regression with constant deviation gives best fit values of $\log(t_c/T_i)= 1.267^{+0.023}_{-0.023}\Delta-2.551^{+0.120}_{-0.127}$ and $\sigma=0.692^{+0.017}_{-0.019}$.  The $\pm$ values show the 95 per cent Highest Posterior Density (HPD) regions). 

To detail the heteroscedasticity of the data, we build two models---one with linearly increasing and another with quadratic increasing standard deviation of stability time with separation.  The response variable in each model is modeled as $y \sim \mathcal{N}(b\Delta+c, b_\sigma \Delta+c_\sigma)$ and $y \sim \mathcal{N}(b\Delta+c,a_\sigma \Delta^2+b_\sigma\Delta+c_\sigma)$, respectively.  The MCMC with the linearly increasing deviation produces a model of stability time where $\log(t_c/T_i)= 1.233^{+0.020}_{-0.019}\Delta-2.382^{+0.091}_{-0.088}$.  These parameters are slightly better constrained than the constant model.  The deviation is given by $\sigma= 0.200^{+0.012}_{-0.012}\Delta-0.396^{+0.054}_{-0.056}$.  The parameters for the constant, linear, and quadratic models are given in Table \ref{tab:mctest}.

\begin{figure}
\begin{center}
\includegraphics[width=\columnwidth]{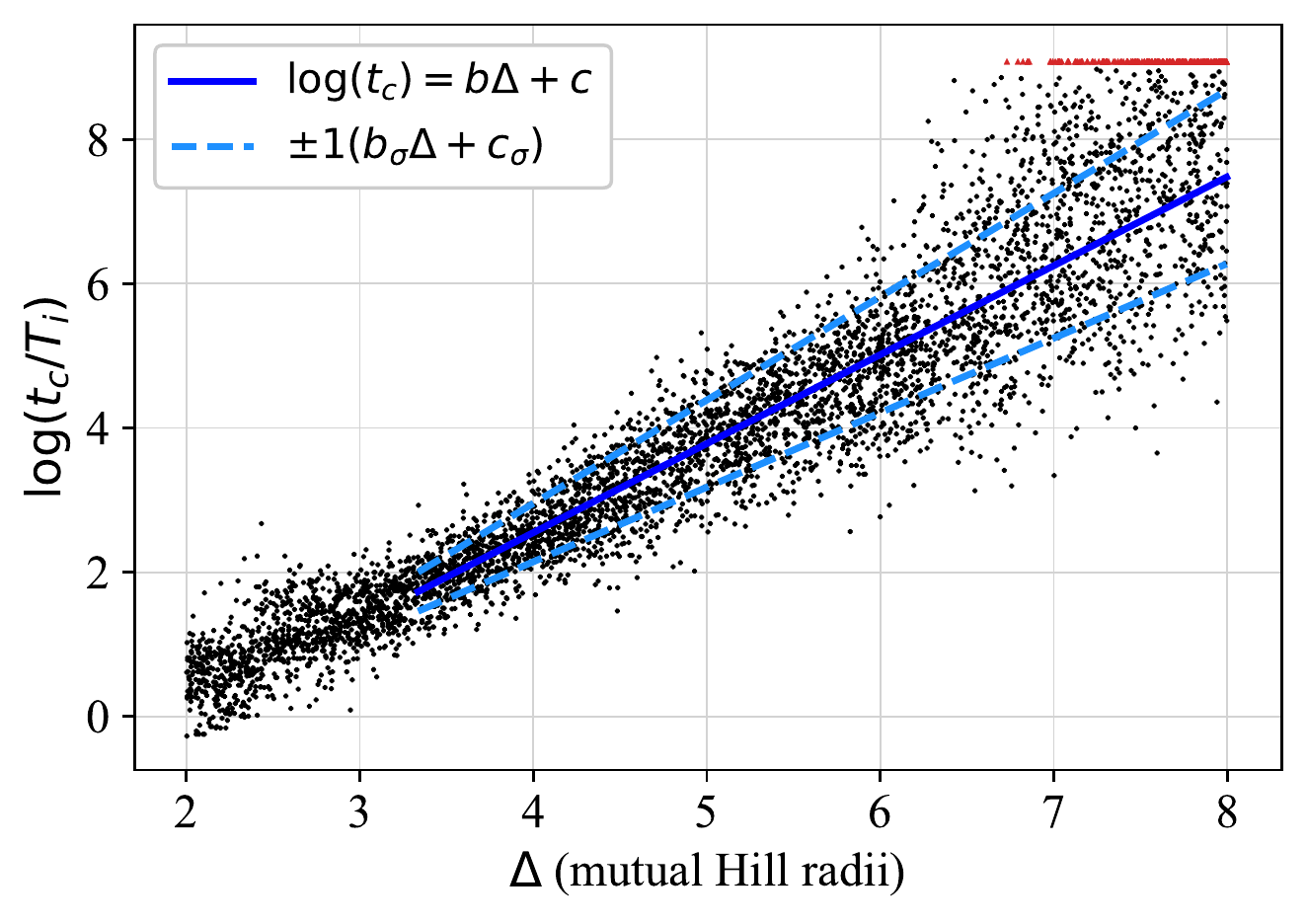}
\includegraphics[width=\columnwidth]{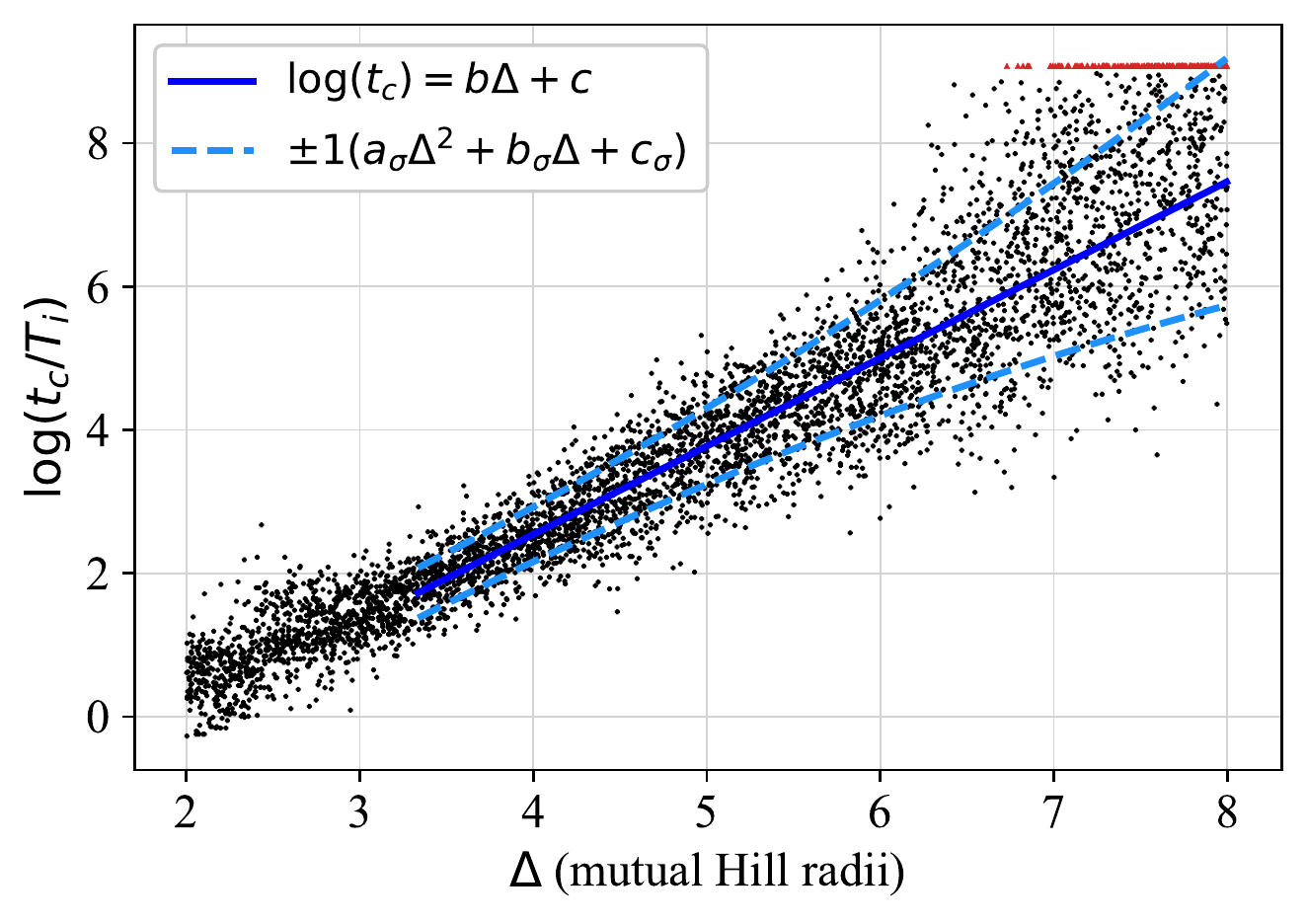}
\end{center}
\caption{Stability time shown for 5,000 systems simulated with $\Delta=[2,8)$ and model linear regression.  \textit{Top} plots the model with linearly increasing standard deviation and \textit{Bottom} plots the quadratic increasing standard deviation.  We only plot the mean values from the posterior distributions for simplicity.  Analysis is done between $\Delta_c<\Delta<7$, but lines are extrapolated here to $\Delta=8$.}
\label{fig:finalchamb}
\end{figure}

We use the mean value of the posteriors for simplicity when plotting.  Fig. \ref{fig:sigma} shows the equations for the standard deviation for our two models.  We compare this to the measured deviation in our data by binning the systems in 0.1 wide bins in $\Delta$ and measuring the bin's deviation in stability time.  Fig.~\ref{fig:finalchamb} shows the models extended past our analysis range.

To compare the goodness-of-fit of our models, we use the Watanabe-Akaike information criterion, also known as the widely applicable information criterion (WAIC) \citep{watanabe}.  In general, the lower the WAIC is the better the model fits the data.  In Table \ref{tab:mctest}, we report the WAIC and the standard error in the criterion for our three Bayesian linear regressions with different models for the standard deviation.  There is a large drop in WAIC when using a non-constant deviation model.  The quadratic model has the lowest WAIC, but it is well within the standard error of the WAIC for the linear model.  Within our analysis range, it remains unclear of how to best model the heteroscedasticity of the stability time.

\begin{table}
	\centering
	\caption{Three models of the data with values for the best-fit line and fits for how the standard deviation in stability time varies with orbital spacing with the average of the HPD region and the Widely-Applicable Information Criteria.\label{tab:mctest}}
	\addtolength{\tabcolsep}{-1pt}
	\begin{tabular}{lccccc} 
		\hline
		Model &            &  b   & c         &      & \\
		 & a$_\sigma$ & b$_\sigma$ & c$_\sigma$ & WAIC & SE\\
		\hline
		Constant  &  & 1.267(23) &  -2.551(124)  &   &   \\
		   &  &  &  0.692(18)  & 6378  &  114 \\
        Linear  &  & 1.233(20) & -2.382(90) &    &  \\
           &  & 0.200(12) & -0.396(55) &  5491  &  94 \\
		Quadratic  &  & 1.230(21) & -2.378(97) &    &   \\
		   & 0.062(14) & -0.408(142) & 1.021(342) &  5415  &  91 \\
		\hline
	\end{tabular}
\end{table}

\subsection{Stability at Large Separations}\label{sec:long}

Our analysis thus far has been bound by the critical separation and $\Delta=7$.  A large portion of systems with separations of $\Delta>7$ have stability times which exceed our maximum integration time.  Previous studies have suggested that the exponential relationship breaks and there is a rapid increase in stability time at large separations where ``large'' is dependent on system properties such as mass and multiplicity \citep{smith,obertas,Petit2020}.  We extrapolate the relationships from the previous section and confirm that systems are more stable than predicted past $\Delta=7$. 

\begin{figure}
\begin{center}
\includegraphics[width=\columnwidth]{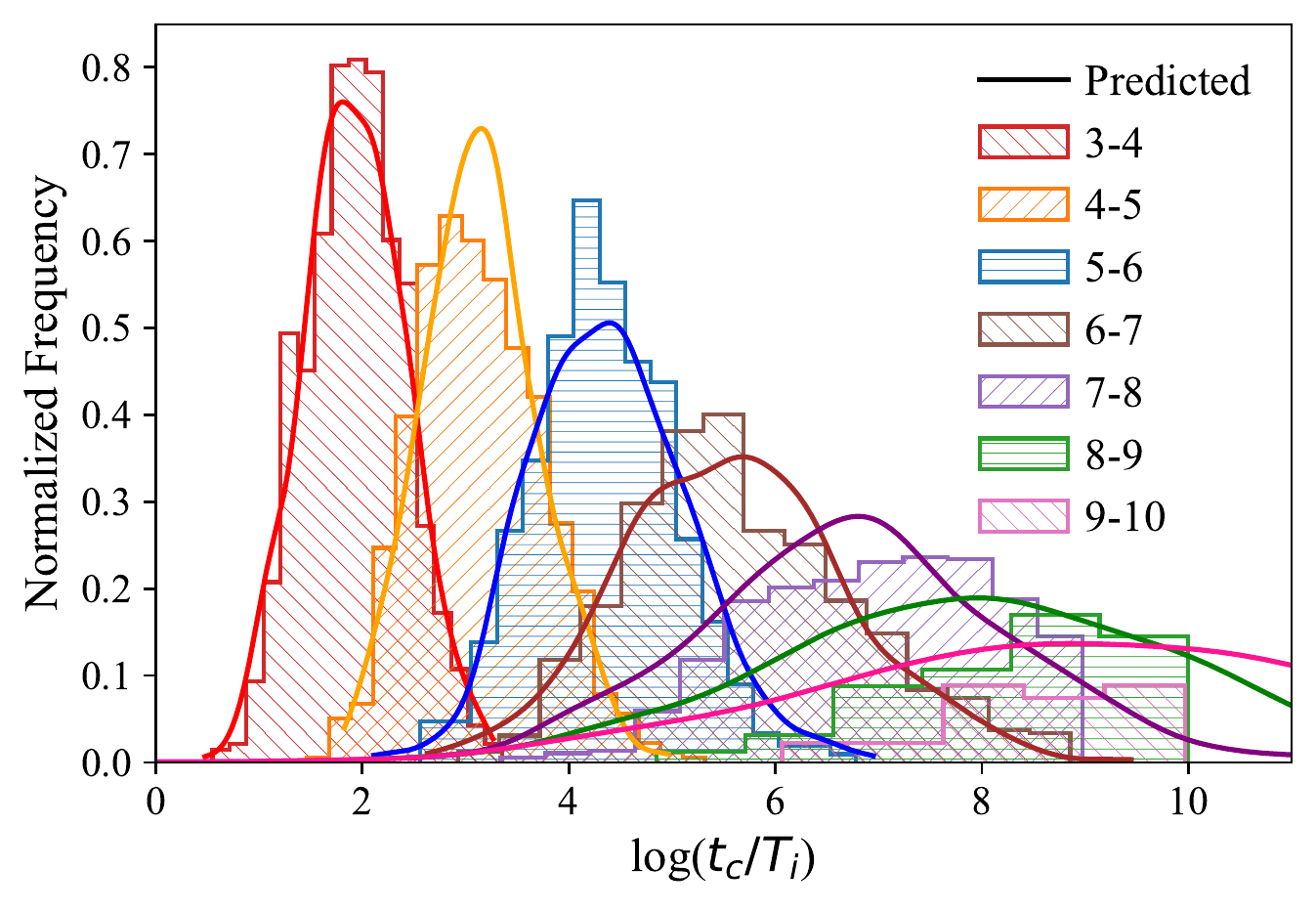}
\end{center}
\caption{Normalized histograms of stability times for the systems in Fig.~\ref{fig:fullchamb} divided into one mutual Hill radii bins which are denoted in the legend (i.e. $\Delta=3$ to 4).  Histograms normalized to the percent with stability time within the maximum integration time.  To determine the ``predicted'' distribution, five million spacings are drawn uniformly between $\Delta=3$ and 8 and each are assigned a stability time chosen using the quadratic deviation model.  These ``predicted'' times are separated into the same bins and are then smoothed with a Gaussian kernel density estimator.  These smoothed distributions fit well for systems with $\Delta<7$, but do not describe stability time at large spacing.  Scott's Rule used for all distributions \citet{scott}.}
\label{fig:dist}
\end{figure}

In Fig.~\ref{fig:dist}, the stability times for the 5000 systems with diverse mass are divided from $\Delta=3$ to 8 into bins with width of 1-$\Delta$.  The 500 systems with an extended integration time of $10^{10}$ orbits are included from $\Delta=8$ to 10.  The skewness of stability time is low for the bins where nearly all systems have stability time within the maximum integration time.  The Fisher-Pearson coefficient of skewness is 0.10, 0.32, and 0.20 for the first three bins; the $\Delta$=6-7 bin is slightly skewed with a coefficient of 0.48.  At small separations, the stability time is distributed log-normally in these large bins in agreement with \citet{chatterjee}, \citet{Rice18}, and \citet{Hussain2020}.  However, at separations wider than $\Delta$=6 the distribution may be more similar to a Rayleigh distribution.

Also in Fig.~\ref{fig:dist}, we show the predicted distributions of stability times from the quadratic deviation model.  These distributions are similar to the histograms up to $\Delta=7$.  The three histograms for $\Delta=7-10$ are shifted to the left with a dearth of systems with short stability times and more with longer stability times.

185 systems have stability times greater than the integration time of $10^9$ orbits for our 5,000 systems with $2<\Delta<8$.  Systems are much more stable than predicted---the linearly increasing standard deviation model predicts 30 systems while the quadratic deviation model predicts 67 systems with stability times over $10^9$ orbits.  At least twice as many systems are stable for over $10^{10}$ orbits for our 500 systems simulated uniformly from $7<\Delta<10$ than expected.  48.2 per cent have stability times over the integration time while the linear and quadratic models predict 14 per cent and 20 per cent, respectively.

There is also a lack of systems with short stability times of less than $10^6$ orbits.  For the 500 systems, 6 per cent have short stability times while the linear model predicts around 10 per cent and the quadratic model predicts 18 per cent.  No systems between $9<\Delta<10$ have short stability time.  Although the linear model predicts only 2 systems in this region should be short, the quadratic model predicts 20 systems.  These results confirm an increase in stability time from the predicted linear fits at large separations.

This increase in stability time at large separations is predicted by \citet{Petit2020}.  They predict that beyond the region of overlapping three planet mean motion resonances there is a rapid increase in the stability time of systems.  An approximate location for the adjacent period ratio between planets for equal-spacing equal-mass systems where this occurs is given by
\begin{equation}\label{eq:maxpratio} 
\frac{P_{i+1}}{P_{i}} = e^{3.48\left(n_r\frac{m}{M}\right)^{1/4}}.
\end{equation}
where $n_r$ is a multiplicative factor for more than 3-body systems.  \citet{Petit2020} finds $n_r$=2.0 works well for the 5-planet systems of \citet{obertas} as there are twice as many resonances in the network, and for 4 planets we use $n_r$=1.5.  For the average mass, 5.5 M$_\oplus$, of these systems the period ratio where the change in stability time occurs is 1.28 which is a separation of 7.4 mutual Hill radii in equal-mass systems. 
 This agrees well with our times.  A 2-sample Kolmogorov-Smirnov test of our simulated and predicted data with $7<\Delta<7.4$ and maximum time of 10$^9$ has a p-value of 0.2 while a test of the simulated and predicted with $7.4<\Delta<8.0$ has a p-value of 10$^{-7}$.

Most studies of stability time limit their computation time by $10^8-10^9$ orbits of the innermost planet, while observed planetary systems with planets within 1 AU often have lifetimes of over one giga-year.  Understanding the stability time at large spacings and how being near strong MMR's causes variability should be of interest to future studies.

\section{Conclusion}\label{sec:conc}

The time until an orbital crossing in a non-resonant multiple planet system can be predicted by the dynamical spacing of the system measured in mutual Hill radii.  Studies that use systems with equally-spaced and equal-mass planets find that the stability time varies from the prediction when systems have period ratios near commensurability with a MMR which causes separation-dependent modulations in the stability time at predictable period ratios.  We disrupt the chains of period ratios by using non-equal masses while keeping the system equally spaced in terms of mutual Hill radii.

Because of the variations in the stability time relationship from MMR, the least-squares regressions used in previous studies \citep{chambers, obertas, wu, gratia} do not account for heteroscedasticity and depend on the range chosen for analysis.  In Fig.~\ref{fig:reschamb} and Fig.~\ref{fig:searth}, our least-squares regression for equal-10 $\mathrm{M_\oplus}$ systems is influenced by the large increase in stability time near $\Delta\approx6$.  This bump is surrounded by the 5:4 and the 4:3 first order MMR between adjacent planets which occur at separations of approximately $\Delta=5.5$ and $\Delta=7$.  We find adding a variation to the masses of the planet's in a system smooths the stability time with spacing relationship over MMR modulations.  

For Earth-mass, we show a variation of 30 per cent is effective in removing structure (Fig.~\ref{fig:varmass} and \ref{fig:structure}).  The mass variation to erase modulations is 10 per cent for 0.1 M$_\oplus$ and 40 per cent for 10 M$_\oplus$.  While systems with higher planet multiplicities have slightly lower stability times and increased clustering of times in the modulations,  the same mass variation is needed to erase modulations in systems of 4 to 20 planets.

The non-equal masses change the period ratio between planets in each system, shown in Fig.~\ref{fig:periodsig}.  The deviation in period ratio is non-linear, but it is approximately a linear effect with $\Delta$ in our analysis region.  By varying the mass of the planets the ratios vary twice as much at $\Delta=7$ than at $\Delta=3.5$.  Since, period ratios near MMRs at larger separations---closer to 2:1---cause a larger decrease in stability time the deviation in period ratios to break the chains of period ratios is larger. 

We further our study by choosing each planet's mass from a uniform distribution between one and ten Earth-masses.  As with normally distributed masses, this prescription smooths out resonance modulations, but the exponential relationship between spacing and stability time remains, Fig.~\ref{fig:fullchamb}.  We find that the relationship in log-time is now best captured by a linear relationship with a linearly or quadratic increasing standard deviation, shown in Fig.~\ref{fig:finalchamb}.  However, this relationship fails to capture the increase in stability times for systems with separations over seven mutual Hill radii.  With a linear model, the standard deviation of stability time becomes one 1.0 dex at $\Delta=6.98$.  At this spacing the time to a close encounter can vary from $10^3$ to $10^9$ orbits of the innermost planet.

The results here suggest that in initially similar systems the time until dynamic instability occurs could be multiple orders of magnitude longer or shorter than predicted from relationships of stability time and spacing.  Two aspects explored in this work, the width of the distribution and stability times at large separations, are possible tests for the analytical explanation in \citet{Petit2020}.

Only a few observed Kepler systems have separations below 8 R$_H$ \citep{pu} and period ratios below 1.4 \citep{steffen, wu}.  Particularly compact systems include Kepler-11 ($\Delta_{min}\approx9.4$), Kepler-223 ($\Delta_{min}\approx8.1$), TOI-216 ($\Delta\approx7.3$), and Kepler-36 ($\Delta\approx4.9$)\footnote{Minimum separation found using NASA Exoplanet Archive, DOI 10.26133/NEA12, as of 11 August 2022} \citep{Lissauer2013,Mills2016,Vissapragada2020}.  The spacings of the TRAPPIST-1 system range from $\Delta\approx 6.57 - 12.8$ and the deviation of mass is 0.42 M$_\oplus$ \citet{Agol2021}; placing it within the context of this work.  However, for the planets of TRAPPIST-1 to survive 1-2$\times 10^{12}$ orbits of the innermost planet \citep{Burgasser2017} the system must have orbital resonances.  TRAPPIST-1, Kepler-223, and TOI-216 have observed orbital resonance \citep{Mills2016,Luger2017,Dawson2021} which stabilize the system \citep{Matsumoto2012, Pichierri2020, Nesvorny2022}.  Improving our understanding of resonant dynamics in multi-planet systems will help constrain the stability of observed compact systems \citep{Tamayo2020}.

The variable stability times found in this work impact planet formation.  During oligarchic growth, large planet embryos have spacings based upon their feeding zones at some multiple of their Hill radii \citep{Kokubo1998,Kokubo2000}.  At the end of planet formation, dynamic instabilities become the dominate form of embryo-embryo interaction \citep{goldreich}.  When systems are not protected by resonances, the timescales investigated in this work control how compact systems evolve to their long-term stable architecture.

\section*{Acknowledgements}

DRR and JHS appreciate A. Yalinewich, C. Petrovich, A. Obertas, D. Tamayo, and A. Petite for stimulating discussions.  We appreciate the constructive feedback of an anonymous referee.  We acknowledge support from the College of Sciences at the University of Nevada, Las Vegas, the Nevada Center for Astrophysics, and NASA grants NNX16AK32G and NNX16AK08G.  All simulations were supported by the Cherry Creek Cluster at UNLV.  We acknowledge that the study resulting in this publication was assisted by a graduate fellowship from The Nevada Space Grant Consortium.

\section*{Data Availability}
Full data, summary data, and plotting tools underlying this article will be shared on request to the corresponding authors.

%%%%%%%%%%%%%%%%%%%%%%%%%%%%%%%%%%%%%%%%%%%%%%%%%%



\bibliographystyle{mnras}
\bibliography{references} % if your bibtex file is called example.bib

% Don't change these lines
\bsp	% typesetting comment
\label{lastpage}
\end{document}